\documentclass[aps,pre,twocolumn,superscriptaddress,floatfix,10pt]{revtex4-1}
\usepackage{graphicx}
\usepackage{dcolumn}
\usepackage{bm}
\usepackage{latexsym}
\usepackage{array,booktabs}
\newcolumntype{L}{@{}>{\kern\tabcolsep}l<{\kern\tabcolsep}}
\usepackage{colortbl}
\usepackage{xcolor}
\usepackage{amsmath}
\begin{document}
\title{Molecular Force Spectroscopy of kinetochore-microtubule attachment {\it in-silico}: mechanical signatures of an unusual catch-bond and collective effects}
\author{Dipanwita Ghanti} 
\affiliation{Department of Physics, Indian
  Institute of Technology Kanpur, 208016, India} 
\author{Shubhadeep Patra} 
\affiliation{ISERC, Visva-Bharati, Santiniketan 731235, India}
\author{Debashish Chowdhury{\footnote{Corresponding author; e-mail: debch@iitk.ac.in}}}
\affiliation{Department of Physics, Indian Institute of Technology
  Kanpur, 208016, India}

\begin{abstract}
Measurement of the life time of attachments formed by a single microtubule (MT) with a single  kinetochore (kt) {\it in-vitro} under force-clamp conditions had earlier revealed a catch-bond-like behavior. In the past the physical origin of this apparently counter-intuitive phenomenon was traced to the nature of the force-dependence of the (de-)polymerization kinetics of the MTs. Here first the same model MT-kt attachment is subjected to external tension that increases linearly with time until rupture occurs. In our {\it force-ramp} experiments {\it in-silico}, the model displays the well known `mechanical signatures' of a catch-bond probed by molecular force spectroscopy. Exploiting this new evidence, we have further strengthened the analogy between MT-kt attachments and common ligand-receptor bonds in spite of the crucial differences in their underlying physical mechanisms. We then extend the formalism to model the stochastic kinetics of an attachment formed by a bundle of multiple parallel microtubules with a single kt considering the effect of rebinding under force-clamp and force-ramp conditions. From numerical studies of the model we predict the trends of variation of the mean life time and mean rupture force with the increasing number of MTs in the bundle. Both the mean life time and the mean rupture force display nontrivial nonlinear dependence on the maximum number of MTs that can attach simultaneously to the same kt. 
\end{abstract}
\maketitle
\section{Introduction}

A mitotic spindle \cite{wittmann01,karsenti01,helmke13} is an example of a self-organized \cite{karsenti08} multi-component molecular machine \cite{frank11} that carries out mitosis \cite{mcintosh12}, i.e., the process of segregation of replicated chromosomes, in eukaryotic cells. These machines are assembled at the right place at the right time and disassemble after serving their biological function.  Even after the mitotic spindle is fully assembled, its size, position, orientation as well as the architecture keep changing dynamically with time, as required for its function. Many of these changes of the spindle are driven by its own components that transduce input energy to generate these forces. Understanding its ``emergent mechanics'' \cite{dumont14}, i.e., how its mechanical 
properties emerge from the complex dynamics, interactions and feedback of its energy-consuming  active building blocks \cite{reber15}, is one of the aims of research on molecular bio-mechanics at the interface of  physics and biology. 

Assembling a mitotic spindle requires formation of molecular joints between specific components. One of the major components of a spindle, that also forms all the key molecular joints in it, is a stiff filament called microtubule (MT) 
\cite{lawson13} each of which has a tubular structure. On the surface of each sister chromatid, that results from DNA 
replication, a proteinous complex called kinetochore (kt) is located \cite{cheeseman14}. During the self-assembling of the spindle each kt attaches with one or more MTs; the actual number varies from one species to another.  In this paper we study kt-MT attachment in a mitotic spindle as an example of a {\it transient} joint in a multi-component molecular machine. This molecular joint plays important roles not only in the morphogenesis \cite{petry16} but also in the subsequent emergent mechanics of the mitotic spindle. 

Force plays all the three roles, namely, input, output and signal, for different components of the same kt-MT attachment \cite{yusko14}. Force exerted by MTs, the key force generators in mitosis, on the kt is essential for proper positioning of the chromosomes in the initial stages of mitosis. Equally important is the opposing tensions exerted by the MTs attached to the two sister chromatids that pull the two sister chromatids apart and away from each other in the late stages of mitosis \cite{mcintosh12}. 

In order to exert force through polymerization/depolymerization, the tip of each MT remains free to polymerize/depolymerize \cite{margolis81} and rapid turnover of its monomeric subunits continues even when the kt-MT attachments remains intact. A kt-MT attachment  would rupture spontaneously, even in the absence of any externally applied tension, by a thermally activated  hopping over a barrier that separates the bound state from the unbound state in the energy landscape \cite{evans98}. Moreover, the kt-MT attachment experiences quite high levels of tension at various stages of mitosis. How the structural integrity of a kt-MT joint is maintained during the entire lifetime of the spindle, in spite of these potentially disrupting tendencies, is one of the wonders of spindle operation.

In this paper we study theoretical models of molecular joints formed by the attachment of $N$ ($N \geq 1$) parallel MTs to a single kt  by treating it as an unusual ``ligand-receptor'' bond. In the words of Martin Karplus \cite{karplus10}, ``the ligand can be as small as an electron, an atom or diatomic molecule and as large as a protein''. In principle, this definition of a ligand can be extended even further to include a MT \cite{lawson13}, whose tubular filamentous structure consists of a hierarchical organization of many proteins. Analogously, the kt, a macromolecular complex hub, is a receptor for a MT. The protocols of `force-clamp' and `force-ramp' that we implement in the computer simulations of our models may be regarded as {\it in-silico} analogs of the corresponding {\it in-vitro} experiments carried out with common chemical ligand-receptor bonds \cite{bizzarri12}. 

In a force-clamp experiment the magnitude of the externally applied tension $F$ on a pre-formed kt-MT attachment is kept fixed (`clamped') and the duration for which the attachment survives before getting fully ruptured is defined as its {\it lifetime}. Similarly, in a ``force-ramp'' protocol \cite{franck10} the magnitude of the tension is increased (`ramped up') with time in a well-defined manner till the bond ruptures at a value of the tension that is identified as the {\it rupture force}. Because of the intrinsically stochastic nature of the process of rupture, both the lifetime and rupture force of a kt-MT joint are random variables that fluctuate from one kt-MT joint to another identical joint. By computing the probability distributions of the lifetime and rupture force and, then, analyzing the data in the light of the analogy with ligand-receptor bonds, we address some fundamental questions on the biomolecular mechanics of the kt-MT joint in a mitotic spindle. 

Our {\it in-silico} studies have been motivated by the {\it in-vitro} biophysical experiments \cite{biggins13,akiyoshi12,sarangapani14} carried out over the last few years using reconstituted kinetochores of budding yeast which happens to be the simplest because each kt can attach with only a single MT \cite{biggins13}.
Under force-clamp conditions, created {\it in-vitro} using optical trap \cite{akiyoshi10}, the mean lifetime of the reconstituted kt-MT attachment of budding yeast was found to increase with increasing tension up to a moderate level beyond which the mean lifetime decreased with further increase of tension. Such nonmonotonic variation of the average lifetime with increasing strength of the pulling force is reminiscent of catch-bonds formed by wide varieties of ligands with their respective receptors \cite{thomas08a,thomas08b,sokurenko08,prezhdo09,chakrabarti17}.

Akiyoshi et al.\cite{akiyoshi10} could account for the catch-bond-like behaviour of the reconstituted kt-MT attachment, as displayed by experimental data, with a 2-state model \cite{bargesov05}. However, this simple model reveals 
neither the structural nor the kinetic origins of this behavior.  Subsequently, a minimal theoretical model was developed by Sharma, Shtylla and Chowdhury
 (from now onwards referred to as SSC model) \cite{sharma14} that explicitly describes the polymerization and depolymerization of the MT. The SSC model reproduced the universally accepted `mechanical signatures of catch bonds' in {\it force-clamp} experiments and elucidated the crucial role of MT kinetics (particularly its force-dependence) that makes this catch-bond unique and unusual. 

In the first part of this paper, we present further evidence in favour of this catch-bond-like behavior by demonstrating that the SSC model also reproduces the well known `mechanical signatures of catch-bonds' in {\it force-ramp} experiments. 
In the second part of this paper we push the analogy with ligand-receptor bonds even further to situations where a bundle of parallel MTs (i.e., multiple ``ligands'') form non-covalent bonds with a single kt (i.e., the ``receptor'').
Studies of the case $N>1$ are important for several reasons. First, it is a natural curiosity because such systems are very common in biological systems. Except unicellular eukaryote budding yeast, cells of most of the organisms, including mammals, have multiple MTs attached with single kt. For example, about 20-40 parallel MTs are bound to each kt in the metaphase spindles of mammalian cells. Analyzing this extended version of the model with $N>1$ under both force-clamp and force-ramp conditions we make new theoretical predictions.

Second, from the perspective of physics, this system provides a unique opportunity to explore collective effects in force generation. Collective force generation by a bundle of polymerizing biofilaments like MTs have  been studied both experimentally as well theoretically (see ref.\cite{bameta17} for a recent overview). Similar fundamental questions on the collective effects of MT bundles in the MT-kt attachments are addressed in this paper. What makes the problem very interesting is that the kinetics of the individual MTs get influenced by others bound to the same kt in spite of the fact that there is no direct interactions between them; the interactions between the MTs are like feedbacks mediated by the kt to which all these MTs are attached. 

At least two physical phenomena add to the complexity of the process of rupture if  $N > 1$. For example, up on detachment of a MT from the kt, the load it experienced before detachment must now be shared by the $n$ (provided $n > 0$) MTs that are still attached to the same kt according to some load-sharing formula. Since, as seen in the special case $N=1$, increasing load on a single MT  does not necessarily destabilize it, the extra load is likely to have a nontrivial effect on the overall stability of the attachment. Moreover, as long as $n \geq 1$, a detached MT can reattach thereby, probably, prolonging the lifetime of the attachment. Do the mean values of the lifetime and rupture force increase simply monotonically, perhaps linearly, with increasing $N$ or is the variation with $N$ more nontrivial? This question is addressed in the second half of this paper

In this paper we also mention the experimental methods that, at least in principle, can test the validity of our theoretical predictions.
It is worth mentioning here that the focus of this work is not on throwing new light on catch-bonds in the usual ligand-receptor systems that have been studied for decades. Instead, our focus is on the tension-induced rupture of the kt-MT attachment, which is an essential transient molecular joint in a functionally important multi-component intracellular molecular machine. However, we discuss this phenomenon from a broader perspective of molecular force spectroscopy of noncovalent ligand-receptor bonds to highlight the crucial differences in spite of the superficial similarities.

\section{SSC model: a brief review}
\begin{figure}[htb]
\center
\includegraphics[angle=0,width=0.5\textwidth]{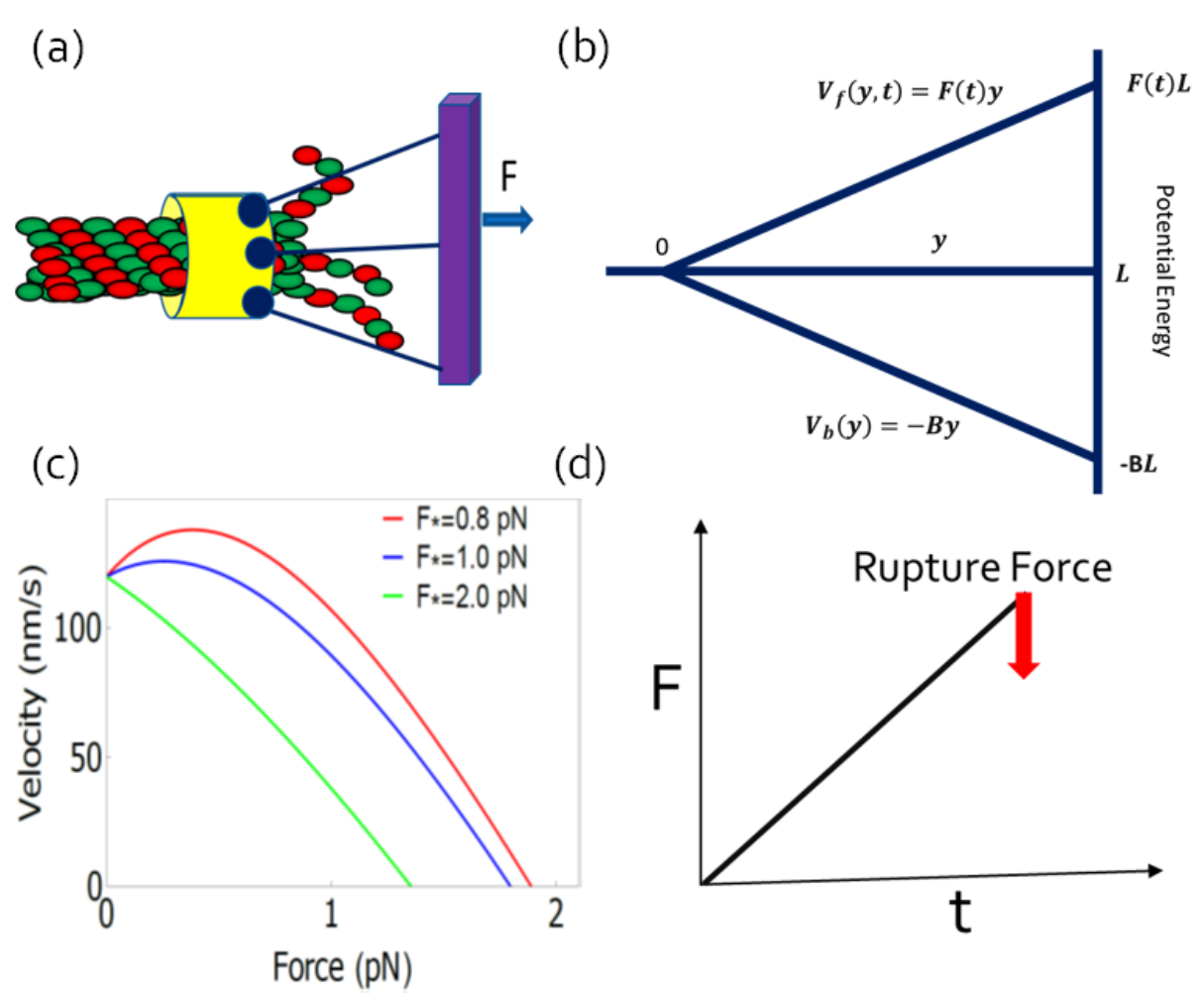}
\caption{(a) A schematic depiction of the kt-MT attachment in the presence of external force. (b) Hypothesized effective potentials $ V_{b}(y) $ and $ V_{f}(y,t) $ are plotted against the instantaneous length of overlap $ y(t) $.
(c)  Net drift velocity $ v(F) $ is plotted against $F$ for three different values 
of $ F_{\star} $.  Parameters used are listed in the table-\ref{table-parameter}.
(d) Linearly increasing force ($ F=at $); different straight lines correspond to different rates of loading. The kt-MT attachment survives the increasing tension up to a certain time and then gets ruptured. } 
\label{fig_model}
\end{figure}

In this section we present a brief summary of the SSC model and review its main results obtained 
earlier under force-clamp condition. This summary will help in motivating the adaptations that are 
appropriate for theoretical analysis of the force-ramp scenario presented in section \ref{sec-rampN1}.
 
The SSC model \cite{sharma14} is a minimal model in the sense that it does not make any 
assumption about the molecular constituents or structure of the kt-MT attachment. It merely 
assumes a cylindrical, effectively ``sleeve-like'', coupler (in the spirit of the Hill sleeve 
model \cite{hill85}) that is coaxial with the MT and has a diameter slightly larger than that of the MT 
(see Fig.\ref{fig_model}(a)). The sleeve may be an abstract representation of the Dam1 ring 
\cite{buttrick11} while the ``rigid rod'', that connects the sleeve with the kinetochore, captures 
the effects of Ndc80 proteins \cite{westermann07,davis07,foley13}. 
No further structural or kinetic details of the Hill model and its later extensions 
\cite{efremov07} (see refs.\cite{asbury11,grishchuk17} for reviews)
have been incorporated in the minimal models of the kt-MT attachments studied here.

Each microtubule is a cylindrical hollow tube with a diameter of approximately 25 nm. 
Globular proteins called $\alpha$ and $\beta$ tubulins form hetero-dimers that assemble
sequentially to form a protofilament. Normally 13 such protofilaments, arranged parallel to 
each other, form a microtubule. The length of each $\alpha$-$\beta$ dimer is about 8 nm. 
However, there is a small offset of about 0.92 nm between the dimers of the neighboring 
protofilaments. SSC model adopts the single protofilament model \cite{anderson13} where 
each MT is viewed as a single protofilament that grows helically with an effective dimer size 
$8/13$ nm. Thus, following the SSC model, throughout this paper, we represent a MT as a 
strictly one-dimensional lattice with the lattice size 8/13 nm. 

In this model the instantaneous overlap between the outer surface of the MT and the inner surface of the 
coaxial cylindrical sleeve is represented by a continuous variable $y(t)$ which is a function 
of time $t$. The total length of the coupler is $L$ so that $0 \leq y(t) \leq L$. Two main postulates 
of this model are as follows \cite{sharma14}:\\
{\it Postulate (a)}: increasing overlap $y$ lowers the energy of the system and that this lowering 
of energy is proportional to $y$ so that the kt-MT interaction potential $V_{b}(y)$ is assumed 
to have the form (see Fig.\ref{fig_model}(b))
\begin{equation}
V_{b}(y)=-By, 
\label{eq_Vb} 
\end{equation}
where $B$ is the constant of proportionality.  Accordingly, the magnitude  of the depth of the potential 
at $ y=L $ is $ BL $. \\
{\it Postulate (b)}: the external force $F$ suppresses the rate of depolymerization $\beta$ of the MT and 
that $\beta$ decreases exponentially with increasing $F$ following  
\begin{equation}
\beta(F)=\beta_{0} exp(-F/F_{\star}), 
\label{eq_betaF}
\end{equation}
where $\beta_{0} $ is depolymerization rate in the absence of any external force and the parameter 
$F_{\star}$ is a characteristic force that determines the sharpness of the decrease of $\beta(F)$ with $F$.\\
The postulate (a) is essentially a limiting case of the Hill model in the sense that the ``roughness'' 
of the interface between the outer surface of the MT and inner surface of the sleeve is neglected 
in the minimal version of the SSC model. The postulate (b) is qualitatively supported by the 
{\it in-vitro} experiments of Franck et al. \cite{franck07}. The decrease of the rate $\beta$  
with the external force $F$ need not be exponential; all the conclusions drawn from the SSC model 
in ref.\cite{sharma14} remain valid as long as the decrease of $\beta$ with increasing $F$ is sufficiently 
sharp. The external tension (using the correct notation for its direction) corresponds to an effective 
potential $V_{f} = F~y$ (see Fig.\ref{fig_model}(b)).

Note that the overlap $y(t)$ can be viewed as the position of a hypothetical Brownian particle in a 
one-dimensional space and subjected to an external potential $V(y) = - B y + F y$. Accordingly,
the kinetics of this model kt-MT attachment can be formulated in terms of a Fokker-Planck (FP)
equation \cite{risken} 
\begin{equation}
 \frac{\partial P(y,t)}{\partial t}=D\frac{\partial ^{2}P(y,t)}{\partial y^{2}}-v(F) \frac{\partial P(y,t)}{\partial y}
\label{eq_FP}
\end{equation}
for the probability density $P(y,t)$, where the net drift velocity 
 \begin{equation}
  v(F)=\frac{B-F}{\Gamma}+(\alpha-\beta(F)){\ell} = \frac{B-F}{\Gamma}+(\alpha-\beta_{0} e^{-F/F_{\star}}){\ell}
 \label{velocity1}
 \end{equation} 
of the hypothetical Brownian particle
involves a phenomenological coefficient $\Gamma$, that characterizes the viscous drag on it, and ${\ell}$ is the increase of the length of MT caused by the addition of each of its subunits. The diffusion constant $D$ in (\ref{eq_FP}) gets contributions from two different physical processes. First, on length scales much longer than ${\ell}$, the stochastic polymerization-depolymerization of a MT can be described in terms of the drift velocity $v$ and an effective diffusion constant \cite{mirny10}
\begin{equation}
D_{MT} = {\ell}^2(\alpha+\beta)/2.
\label{eq-mirnyD}
\end{equation} 
even when the MT tip is not attached, or tethered, to any surface. The second contribution that exists even in the absence of polymerization-depolymerization of the MT is the diffusive motion of the kinetochore plate itself \cite{joglekar02}.  Considering ${\ell}=8/13$ nm, $\alpha$=30 $s^{-1}$ and $\beta << \alpha$, the effective diffusion constant $D_{MT}$ is approximately 5 nm$^2$/s. Even if one includes the maximum possible value of $\beta(F)$, i.e., $\beta_{0}=350$s$^{-1}$ \cite{hill85,joglekar02,shtylla11,waters96} in (\ref{eq-mirnyD}), the effective diffusion constant $D_{MT}$ increases to about $70$ nm$^{2}$/s which is still about an order of magnitude smaller than the contribution coming from the diffusional movement of the kinetochore plate which is typically 700 nm$^{2}$/s \cite{hill85,joglekar02,shtylla11}.  Therefore, throughout this paper, we assume the diffusion constant $D$ to be independent of the external tension $F$.

The FP equation (\ref{eq_FP}) can also be re-cast as an equation of continuity  
\begin{equation}
\frac{\partial P(y,t)}{\partial t}=-\frac{\partial J(y,t)}{\partial y}
 \label{eq-continuity}
\end{equation} 
for the probability density $P(y,t)$ with the probability current density 
\begin{eqnarray}
 J(y,t) &=& -D\biggl[\frac{\partial P(y,t)}{\partial y}-\frac{v(F)}{D}P(y,t)\biggr] \nonumber \\ 
&=&-D\biggl[\frac{\partial P(y,t)}{\partial y}+\frac{U'(y)}{k_{B}T}P(y,t)\biggr].
 \label{flux}
\end{eqnarray}
where $ U'(y)=d\tilde{U}(y)/dy $ and effective potential $ \tilde{U}(y) $ is given by 
\begin{equation}
 \frac{\tilde{U}(y)}{k_{B}T}=\biggl[\frac{F-B}{k_{B}T}+{\ell}\frac{(\beta(F) -\alpha)}{D}\biggr]y.
 \label{eq-effective-pot}
\end{equation}
Note that the terms involving $F$ and $B$ in eq.(\ref{eq-effective-pot}) are of energetic 
origin whereas the remaining two terms involving $\alpha$ and $\beta$ are of kinetic origin.

The attachment survives as long as 
$y$ remains non-zero; the rupture of the attachment is identified with the attainment of the 
value $y=0$ {\it for the first time}.  For the calculation of the lifetime of the attachment a unique 
initial condition is required. In ref.\cite{sharma14} the authors  assumed that initially (i.e., at 
time $t=0$) the MT is fully inserted into the sleeve, i.e., 
\begin{equation}
y(t=0) = L ~~~~~~({\rm initial ~condition}).
\label{eq-ini}
\end{equation}
Since the MT is not allowed to penetrate the kinetochore plate, the overlap $y$ cannot exceed $L$. 
This physical condition is captured mathematically by imposing the {\it reflecting} boundary 
condition 
\begin{equation}
J(y,t)\vert_{y=L}=0. 
\label{eq-bc1}
\end{equation} 
An absorbing boundary condition 
\begin{equation}
P(y,t)\vert_{y=0}=0 
\label{eq-bc2}
\end{equation}
is imposed at $y=0$ for the calculation of the life times. 
In terms of the hypothetical Brownian particle, 
the FP equation for $y(t)$ can be viewed as that for the position of a hypothetical Brownian 
particle, subjected to an external potential $V(y) = - B y + F y$, in a one dimensional 
space with a reflecting boundary at $y=L$, an absorbing boundary at $y=0$ and the initial 
condition $y(t=0)=L$.

Starting from the initial condition $y=L$, the time taken by the kt-MT attachment to attain vanishing 
overlap ($y=0$) {\it for the first time} was  identified as the life time of the attachment.
Thus, the calculation of the lifetime is essentially that of a {\it first passage time} for a hypothetical 
Brownian particle: the time it takes to reach $y=0$ for the first time starting from $y=L$ at $t=0$.
This lifetime fluctuates from one kt-MT attachment to another; the distribution of the lifetime contains 
all the statistical information.

In ref.\cite{sharma14} the authors calculated the exact distribution of the lifetimes analytically 
in the Laplace space and hence the mean lifetime $<t>$  to be  
\begin{equation}
   <t>=\frac{D}{v^{2}(F)}\biggl[e^{v(F)L/D}-1\biggr]-\frac{L}{v(F)}
 \label{eq_timeavg}
 \end{equation}

\begin{figure}[htb]
\center
\includegraphics[angle=0,width=0.5\textwidth]{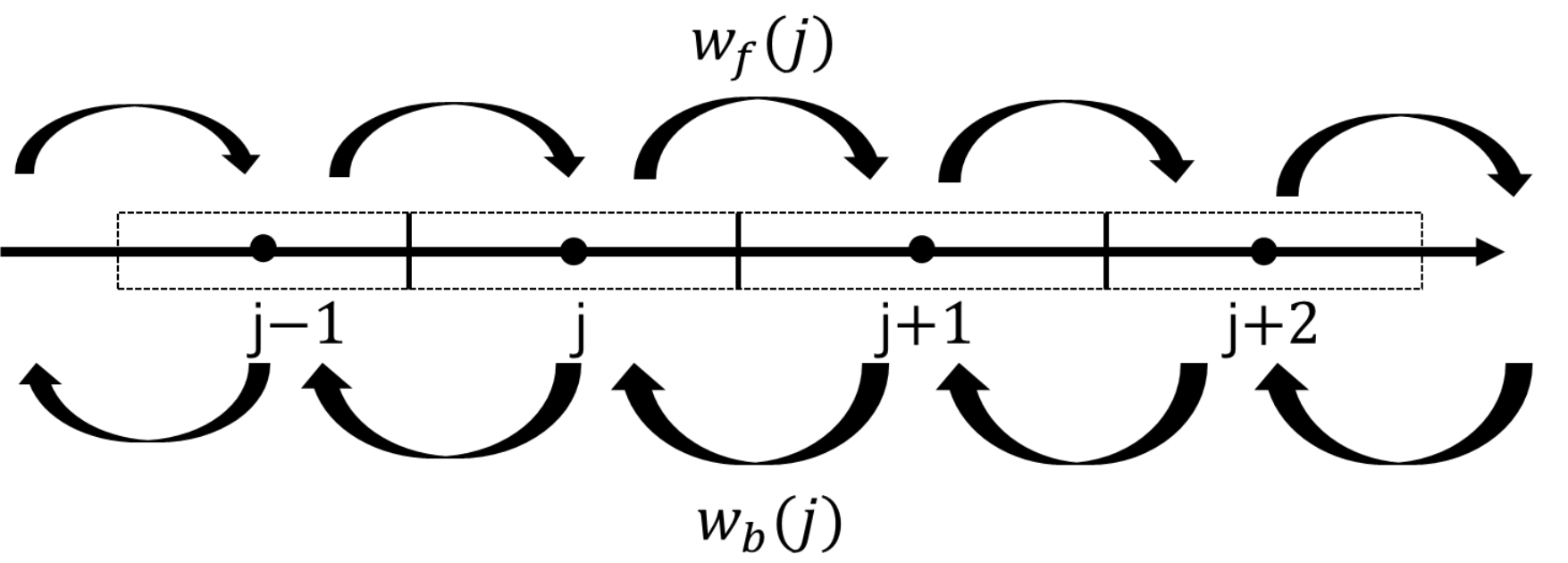}
\caption{Using WPE method, continuous 1-D space of length $ L $ is discretized. In this discretized space MT tip is moving either in the forward or in the backward direction using transition rates $ w_{f}(j) $ and $ w_{b}(j) $.}
\label{fig_wfwb}
\end{figure}

For the convenience of numerical computation of the distribution of the lifetimes by computer 
simulation, the SSC model was discretized in ref.\cite{sharma14} following prescriptions 
proposed earlier by Wang, Peskin and Elston (WPE) \cite{wang03,wang07}. 
WPE method is a numerical algorithm in which a FP equation
is discretized into a discrete Markovian jump process by finite differencing of the FP equation.
Following WPE, space was discretized into $M$ cells, each of length $h = L/M$ and the 
continuous effective potential $\tilde{U}(y)$ was replaced by its discrete counterpart 
\begin{equation}
\frac{\tilde{U}_{j}}{k_BT}=\biggl[\frac{(F-B)}{k_{B}T}+\ell\frac{\biggl(\beta_{0}e^{-F/F_*}-\alpha\biggr)}{D}\biggr] y_{j}
\label{eq-discreteU}
\end{equation}
where $y_j$ denotes the position of the center of the $j$-th cell. In the discrete formulation, 
instead of a FP equation, a master equation describes the kinetics of the system in terms of 
discrete jumps of the hypothetical Brownian particle from the center of a cell to that of its 
adjacent cells, either in the forward or in the backward direction. The rates of forward and 
backward jumps $\omega_{f}(j)$ and $\omega_{b}(j)$ on the discretized lattice were given by 
\cite{wang03} 
\begin{eqnarray}
\omega_{f}(j)&=&\frac{D}{h^2}\frac{-\frac{\delta \tilde{U}_{j}}{k_{B}T}}{\exp\biggl(-\frac{\delta \tilde{U}_{j}}{k_{B}T}\biggr)-1} 
= \dfrac{1}{h}\dfrac{\dfrac{B-F}{\Gamma}+\ell(\alpha-\beta)}{\exp\biggl(-\dfrac{\delta \tilde{U_{j}}}{k_{B}T}\biggr)-1}\nonumber \\
\label{eq_wf}
\end{eqnarray} 
\begin{eqnarray}
\omega_{b}(j)&=&\frac{D}{h^2}\frac{\frac{\delta \tilde{U}_{j}}{k_{B}T}}{\exp\biggl(\frac{\delta \tilde{U}_{j}}{k_{B}T}\biggr)-1} 
= \dfrac{1}{h}\dfrac{\dfrac{F-B}{\Gamma}+\ell(\beta-\alpha)}{\exp\biggl(\dfrac{\delta \tilde{U_{j}}}{k_{B}T}\biggr)-1}
\label{eq_wb} \nonumber \\
\end{eqnarray}
where 
\begin{equation}
\delta \tilde{U}_{j}= \tilde{U}_{j+1}-\tilde{U}_{j}
\end{equation} 
Excellent agreement between the results derived from the analytical theory and computer simulations 
was reported in ref.\cite{sharma14}. 

\section{Force-clamp: dependence of life time for N=1 on initial conditions}

In ref.\cite{sharma14}, where the SSC model was presented, the authors reported the results for the model only under force-clamp conditions. However, as summarized in the preceding section, the lifetimes of the attachments  were calculated beginning always with the unique initial condition $y(t=0)=L$. In order to test whether the conclusions drawn in ref.\cite{sharma14} are sensitive to the choice of the initial condition, we have now carried 
out a detailed investigation of the distribution of the lifetimes with  two different types of initial conditions. 

\begin{figure}[htb]
\center
\includegraphics[angle=0,width=0.5\textwidth]{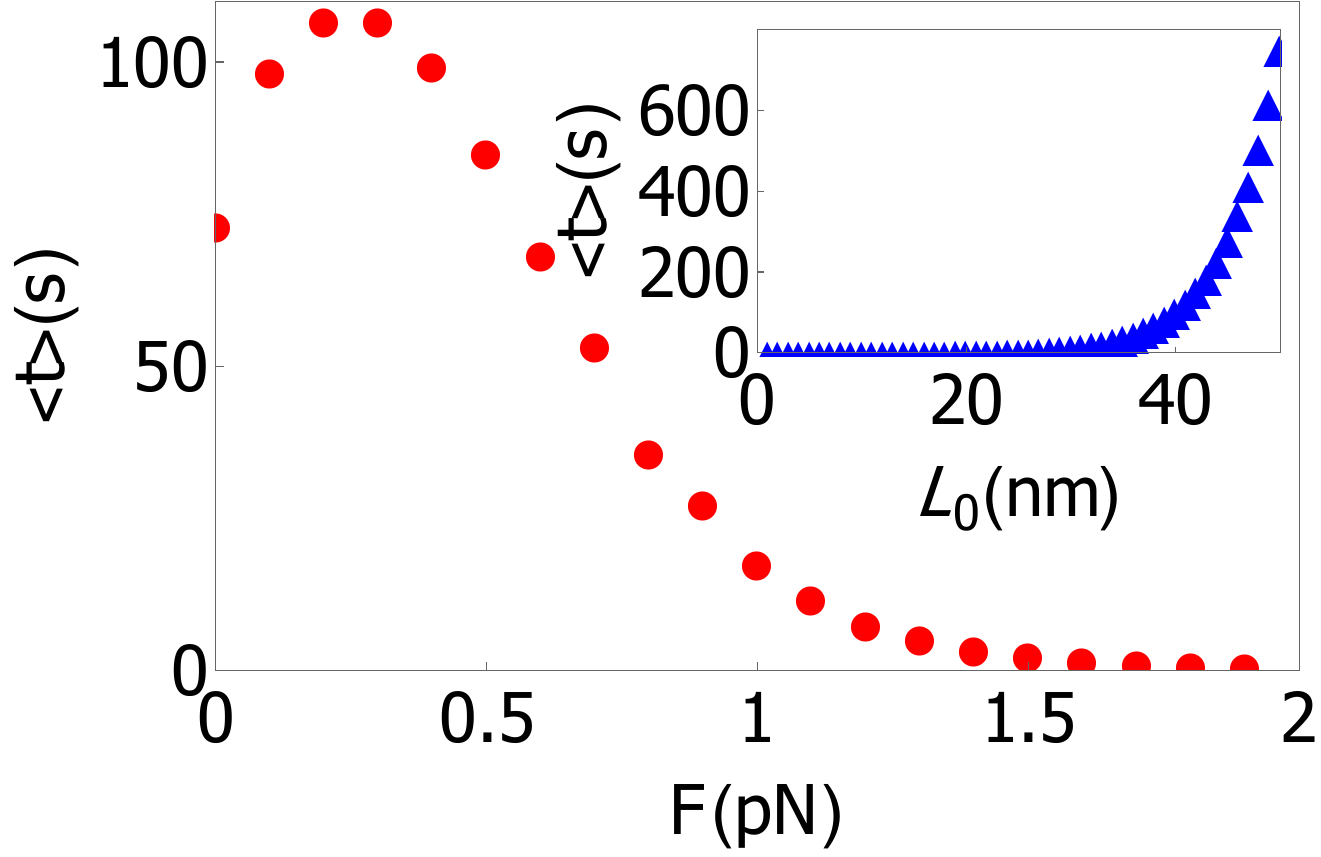}
\caption{Mean lifetime, under force-clamp condition, plotted against external force $ F $ for randomly chosen initial position of the coupler. In the inset mean lifetime is plotted by varying the initial position of the coupler $ L_{0} $ for a fixed force $ F=0.5 $pN. The numerical values of all the other parameters used in the simulation are listed in the table \ref{table-parameter} except, $N=1$, $ F_{*}=1 $pN and $ B=2 $pN.}
\label{fig_initial cond}
\end{figure}

In one of these, any integer lying between $1$ and $L$ is chosen, with equal probability, to be the initial value of the overlap $y$. The mean lifetime is obtained by averaging the lifetimes over a sufficiently large number of samples each with a randomly chosen initial condition. The results of these computations are plotted in Fig.\ref{fig_initial cond}. The ``catch-bond" behavior is still observed. 

In the second type, the lifetimes are first calculated for a {\it fixed} initial condition $y(t=0)=L_{0}$ ($1 \leq L_{0} \leq L$) and then these lifetimes are averaged over a large number of samples, all for the same initial overlap $L_{0}$, getting the mean life time  $<\tau>(L_{0})$ corresponding to the fixed $L_{0}$. The process is then repeated for several different values of $L_{0}$ to get $<\tau>$ as a function of $L_{0}$ ($0 \leq L_{0} \leq L$). The results of these computations are plotted in the inset of Fig.\ref{fig_initial cond}. 

As expected on physical grounds, the mean lifetime of the attachment increases with increasing initial overlap $L_{0}$, attaining its largest value $(\approx 758s)$ for $L_{0}=L$ i.e., where the MT is initially fully inserted into the coupler. Note that the mean lifetime corresponding to $L_{0}=L$ is about seven times that of the attachments where the initial overlap is selected at random. Such lower values of $<\tau>$ for random initial overlaps is expected on the physical grounds that in several initial configurations the MT begins with an initial overlap $y(t=0) < L$ and, hence, expected to rupture sooner that those with initial overlap $y(t=0)=L$.

\noindent $\bullet$ {Physical origin of the catch-bond-like behavior}

The external pulling force $F$ has two opposite effects on the MT. From the expression (\ref{velocity1})
for the net drift velocity $v(F)$ we see that, on the one hand, the MT is bodily pulled out of the coupler 
by $F$. On the other hand, because of the suppression of the depolymerization by the external pull $F$, 
if the depolymerization rate $\beta$ falls  below that of polymerization the tip of the MT exhibits a net 
growth. Moreover, if the suppression of depolymerization is so strong that the net rate of tip growth into 
the coupler (increase in $y$) can more than compensate the rate of bodily exit of the MT from the coupler 
(decrease of $y$) the growing MT tip moves deeper inside the coupler (resulting in the net increase of 
$y$) when subjected to  external tension. Such an increase of $y$ (indicated by increase of $v(F)$ in 
Fig.\ref{fig_model}(c)), instead of the naively expected decrease, upon application of $F$ would be 
interpreted as an effective increase of the stability of the kt-MT attachment with increasing strength of 
the applied force $F$. However, as the strength of $F$ increases, $\beta(F)$ gradually saturates. Since 
$\beta$ practically stops decreasing further with the further increase of $F$ the bodily movement of the 
MT out of the coupler at higher values of $F$ can no longer be compensated by the tip growth into the 
coupler; the net decrease of $y$ (indicated by decrease of $v(F)$ in Fig.\ref{fig_model}(c)) with further 
increase of $F$ in this regime manifests as decrease in the stability of the kt-MT attachment. However, 
monotonic decrease of $v(F)$ with increasing $F$ seen in Fig.\ref{fig_model}(c)) results for larger values 
of $F_{\star}$ because of weak suppression of depolymerization by the external tension.

The physical scenario that emerges from the above interpretation of the dependence of $v(F)$ on 
$F_{\star}$ is also consistent with the expression (\ref{eq_timeavg}) for the mean lifetime $<t>$ 
where $v(F)L$ acts like an effective barrier height. For small enough $F_{\star}$, the nonmonotonic 
variation of $v(F)$ with $F$ manifests as a nonmonotonic variation of the barrier height $v(F) L$, 
resulting in a nonmonotonic variation of the mean lifetime $<t>$ with $F$ which has been interpreted 
as a catch-bond-like behavior. In contrast, for sufficiently large $F_{\star}$ the monotonic decrease 
of $v(F)$ with $F$ results in a monotonic decrease of the effective barrier height  $v(F)L$ which, 
in turn, causes the monotonic decrease of the mean lifetime $<t>$ with $F$ that has been interpreted 
as a slip-bond-like behavior. 
Thus, to summarize, whether the attachment behaves like a catch bond or a slip bond depends crucially 
on the magnitude of $F_{\star}$, which determines the extent of suppression of depolymerization for a 
given $F$, i.e., how sharply the depolymerization rate $\beta(F)$ falls with increasing tension $F$.

\section{Rupture of kt-MT attachment under ramp force for N=1}
\label{sec-rampN1}

In ref.\cite{sharma14} the external tension $F$ was assumed to be independent of time $t$; 
this condition corresponds to a ``force-clamp'' situation in the experiments. In this section 
the time-dependent external tension $F(t)$ is assumed to increase according to a well defined 
protocol; this corresponds to a ``force-ramp'' in experiment (see Fig.\ref{fig_model}(d)). 
We adopt the  postulates (a) and (b) of the SSC model. For the sake of simplicity, we assume 
a linear ramp  force, namely, $ F(t)=a~t $ where $ a $ is the loading rate. The instantaneous 
external tension $F(t)$ can be derived from the corresponding instantaneous potential landscape, 
$ V_{f}(y,t)=F(t)y $ . The effective potentials $ V_{b}(y) $ and  $ V_{f}(y,t) $ at an arbitrary 
instant of time are plotted in Fig.\ref{fig_model}(b). Net instantaneous potential $V(y,t)$ felt 
by the kinetochore is $ V(y,t)= V_{b}(y)+V_{f}(y,t) $.

For the theoretical treatment of the kt-MT attachment subjected to a ramp force $F(t)$, 
we adapt the corresponding theory for ligand-receptor bond rupture, developed originally by 
Bell \cite{bell78} and subsequently extended by Evans and Ritchie \cite{evans97} and by Evans and 
Williams \cite{evans02} (see also the reviews in refs.\cite{evans01,friddle12,arya16}). 
In the presence of a given tension $F$, let $k_{off}(F)$ be the rate (i.e., probability per unit time) 
of unbinding of a MT from the kt. Because of the specific choice of the initial condition $y(t=0)=L$ 
and the absorbing boundary condition at $y=0$, no rebinding of the MT is possible and, therefore, 
rebinding rate remains $k_{on}(F)=0$ throughout this section. 

Denoting the probability that $y\neq 0$ (i.e., MT is attached to the kt) at time $t$ by the 
symbol $P_{on}(t)$, the equation governing the time evolution of $P_{on}(t)$ is 
\begin{equation}
\frac{dP_{on}(t)}{dt} = - k_{off}(F) P_{on}(t).
\label{eq-Pont}
\end{equation}
Hence, in terms of $k_{off}(F)$, the survival probability $S(t)$ of the attachment (i.e., the 
probability that the hypothetical Brownian particle has not reached $y=0$ before time $t$) can be 
expressed as  \cite{friddle12}
\begin{equation}
 S(t)=exp\biggl[-\int_{0}^{t}k_{off}(F(t^{'}))dt^{'}\biggr]
 \label{eq_ramp_surv}
\end{equation}
Moreover, in terms of $k_{off}(F)$ the probability density $\rho_{fp}(F)$ of the rupture forces 
is expressed as \cite{friddle12}
 \begin{equation}
  \rho_{fp}(F)=\frac{k_{off}(F)}{a}\biggl[exp\biggl(-\frac{1}{a}\int_{0}^{F}k_{off}(F')dF'\biggr)\biggr]
 \label{eq_ruptureforce}
 \end{equation} 
  Mean Rupture force is given by \cite{friddle12}
  \begin{equation}
   <F>=\int_{0}^{\infty} F~ \rho_{fp}(F) ~dF
 \label{eq_meanF}
 \end{equation}
Thus, for the calculation of $S(t)$ and $\rho_{fp}(F)$ the analytical expression for $k_{off}(F)$ 
is required. For $k_{off}(F)$ we use the expression for the inverse of the average lifetime of  a 
single kt-MT attachment in the SSC model, reported in  ref.\cite{sharma14}, namely, 
\begin{equation}
   k_{off}(F)=\frac{1}{<t>}=\frac{v^{2}(F)}{D(e^{v(F)L/D}-1)-Lv(F)}
 \label{eq_koff}
 \end{equation} 
where the expression $v(F)$ is given by Eq.(\ref{velocity1}).
The expression (\ref{eq_koff}) was derived under force-clamp condition and, therefore, strictly valid 
when the force does not vary at all with time. Use of this expression for $k_{off}(F)$ in the calculation 
of $S(t)$ and $\rho_{fp}(F)$ is based on the assumption that the expression (\ref{eq_koff}) is a good 
approximation even when the tension varies with time. Obviously, the deviations of $k_{off}(F)$ from 
this expression in force-clamp conditions is expected to be insignificant provided the rate of increase 
of $F$ is sufficiently small. 
Substituting Eq.(\ref{eq_koff}) into the Eqs.(\ref{eq_ramp_surv}) and  (\ref{eq_ruptureforce}) we get, 
respectively, the survival probability $S(t)$ and the rupture force density $\rho_{fp}(F)$ by numerically 
evaluating the respective integrals.

Thus, the theoretical results for the case $N=1$ have been derived from numerical integrations of eqns.(\ref{eq_ramp_surv})-(\ref{eq_ruptureforce}) which have been plotted throughout this section by lines.
For computer simulation of the model, we discretize the FP equation of the SSC 
model following WPE prescription \cite{wang03,wang07} as explained above \cite{sharma14}. 
Instead of a constant force, a time-dependent external force $ F= a t $ is imposed. Carrying 
out computer simulations of this discretized version of the model we directly compute the survival 
probability $S(t)$ and the distribution $\rho_{fp}(F)$ of the rupture forces. Throughout this section, 
discrete symbols have been used to plot the data obtained from computer simulations of the discretized 
model. Parameter values that we used for numerical calculations are listed in table \ref{table-parameter}.

 \begin{table}[t]
\centering
\begin{tabular}{l*{2}{c}r}
Parameter             & Values\\
\hline
Inter-space between MT binding site  $ l $ \cite{hill85,joglekar02,shtylla11}  & 8/13 $ nm $  \\
Total length of coupler $ L $ \cite{gonen12,bloom08,johnson10,salmon06} & 50 $ nm $\\
Polymerization rate $ \alpha $ \cite{hill85,joglekar02,shtylla11,waters96} &  30 $ s^{-1} $ \\
Maximum Depolymerization rate $ \beta_{0} $ \cite{hill85,joglekar02,shtylla11,waters96}  & 350 $ s^{-1} $ \\
Characteristic force of Depolymerization $ F_{\star}$ \cite{sharma14} &  0.8 $ pN $ \\
Attractive force between kt-MT $ B $ \cite{sharma14} &  1.9 $ pN $  \\
Diffusion constant $ D $ \cite{hill85,joglekar02,shtylla11} &  700 $ nm^{2}s^{-1} $    \\
Viscous drag coefficient $ \Gamma $ \cite{hill85,joglekar02,shtylla11,marshall01}  &  6$ pN s ~\mu {\bf m}^{-1} $ \\
\end{tabular}
\caption{Values of the parameters for kt-MT system }
\label{table-parameter}
\end{table}

 
\begin{widetext}

\begin{figure}[htb]
\includegraphics[angle=-0,width=0.8\columnwidth]{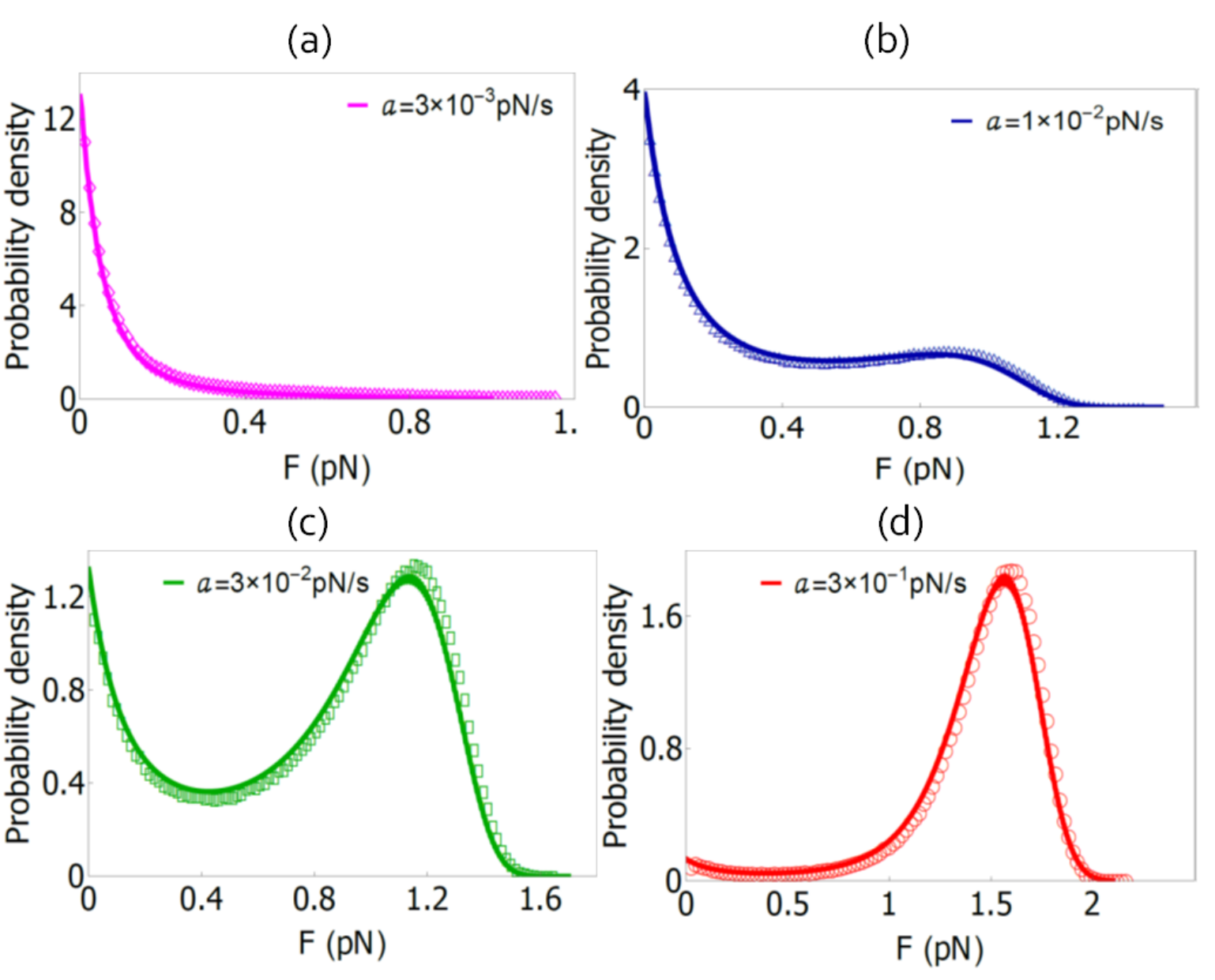}\\
\caption{Probability density of rupture force of the kt-MT attachment with $N=1$, under force-ramp condition,  for four different 
loading rates, namely, (a) $ a=3\times 10^{-4} pNs^{-1} $ (violet rhombus), (b) $ a=1\times 10^{-3} pNs^{-1} $
 (green square), (c) $  a=3\times 10^{-3} pNs^{-1} $(blue triangle) and (d) $  a=3\times 10^{-2} pNs^{-1} $
 (red circle) are plotted. The continuous curves have been plotted by numerical integration 
 of the Eq.(\ref{eq_ruptureforce}) whereas the discrete data points have been obtained from 
 computer simulations of the discretized version of the same model.  Numerical values of all the
 other parameters are listed in table-\ref{table-parameter}.} 
\label{fig_1MT ramp1}
\end{figure} 

\end{widetext}

\begin{figure}[htb]

\includegraphics[angle=-0,width=0.95\columnwidth]{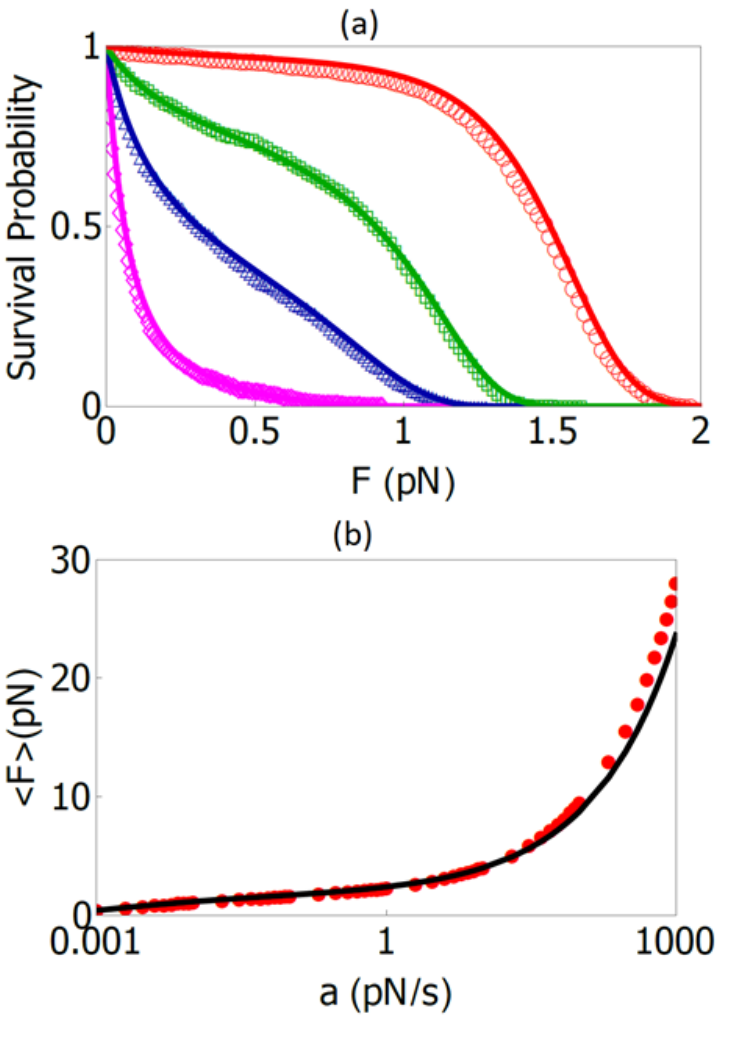}
\vspace{.5cm}
\caption{(a)Survival probability for different loading rates; the continuous curves in (a) have been plotted by numerical integration  of the Eq.(\ref{eq_ramp_surv}).
 (b) Mean rupture force is plotted against the logarithm of the loading rate; the continuous black line in (b) has been plotted by numerical integration of the Eq.(\ref{eq_meanF}).
 The same symbols in (a) and Fig.\ref{fig_1MT ramp1} correspond to the 
 same set of values of the model parameters. Numerical values of all the other parameters are listed in table-\ref{table-parameter}.} 
\label{fig_1MT ramp2}
\end{figure} 


 In the Fig.\ref{fig_1MT ramp1} the rupture force distribution obtained from numerical integration 
 of the eqns.(\ref{eq_ramp_surv})-(\ref{eq_ruptureforce}) of the continuum theory and those obtained 
 from computer simulation of the discretized model are plotted for four different loading rates. 
 At loading rates as low as  $a=3\times10^{-4}pNs^{-1}$ (violet), the most probable rupture force is 
vanishingly small. At such slow loading rates the rupture of the attachment is mostly spontaneous 
dissociation caused by thermal fluctuation and is very rarely driven by the applied tension. However, 
as the  loading rate increases a second peak at a non-zero value of the force begins to emerge. 
At moderate loading rates like $ a=1\times10^{-3}pNs^{-1} $ (blue line and triangle ) and 
$ a=3\times10^{-3}pNs^{-1} $ (green line and square)), a large fraction of the kt-MT attachments 
survive upto a high force before getting ruptured while another significant fraction of the attachments 
still dissociate at a vanishingly small force. But, at sufficiently high rates of loading, for example at 
$ a=3\times10^{-2}pNs^{-1} $ (red),  an overwhelmingly large fraction survives up to a high force while 
very few attachment get ruptured by very weak forces.

In the Fig.\ref{fig_1MT ramp2}(a) the survival probabilities are plotted at  the same loading 
rates for which the rupture force distributions have been plotted in Fig.\ref{fig_1MT ramp1}.      
At very high loading rates the probability of survival remains high, 
and practically unaffected  by the applied force, upto quite high values of the force and, 
accordingly, the most probable rupture force is also expected to be high. In contrast, sharp drop 
in the survival probability with increasing force is also reflected in the vanishingly small most 
probable rupture force at very low loading rates. 

In the Fig.\ref{fig_1MT ramp2}(b) we have plotted 
mean rupture force as a function of loading rate. Mean rupture force increases with increasing loading rate.
The increase of mean and most probable rupture force with increasing loading rate is also 
observed in case of common ligand-receptor attachments \cite{friddle12}; it follows from the 
mathematical form of the equation 
\begin{equation}
\frac{dP_{on}(F)}{dF} = - \frac{1}{a} k_{off}(F) P_{on}(F) 
\label{eq-PonF} 
\end{equation}
which is nothing but the equation (\ref{eq-Pont}) expressed in terms of force $F$ rather than 
time $t$. Eqn.(\ref{eq-PonF}) implies that the rate of decay of the bound state of the bond 
is inversely proportional to the loading rate $a$. Consequently, the kt-MT attachment  
persists up to higher values of force when subjected to faster loading rates.  

The continuous black line in Fig.\ref{fig_1MT ramp2}(b) has been obtained using Eq.(\ref{eq_meanF}).  
As the loading rate exceeds about $100$ pN/s, the black line begins to deviate from the corresponding 
data points obtained from simulations. This increasing deviation indicates increasing failure of the 
approximation made by substituting the force-clamp values of $k_{off}(F)$  for evaluating the integrals 
in Eq.\ref{eq_ruptureforce}. However, surprisingly, even at ten times faster loading rates the error made by 
this approximation is within about $20 \%$.

Irrespective of the actual loading rate, a slip bond exhibits a single peak at $F=F_{mp}$ in the rupture force distribution $\rho_{fp}(F)$ at a given rate of loading. In this case,  the most probable rupture force $F_{mp} \to 0$ corresponding to $a \to 0$ and $F_{mp}$ increases with loading rate $a$. The trend of variation of $\rho_{fp}(F)$ with the loading rate $a$ is qualitatively different in case of catch bonds. For the latter, at sufficiently low values of $a$, the distribution $\rho_{fp}(F)$ exhibits a high peak at $F=0$ and a much lower peak at a larger nonzero value of $F$ while $\rho_{fp}(F)$ remains very small over a wide range of $F$ in between these two peaks. With the increase of the loading rate $a$ the second peak at the non-zero $F$ increases in height while a concomitant lowering of the peak at $F=0$ occurs. The occurrence of two peaks in $\rho_{fp}(F)$ for a given $a$ is regarded as the `mechanical signature' of catch bond in force-ramp experiments \cite{evans04,thomas08a,thomas08b}. 

The shape of $\rho_{mp}(F)$ plotted in Fig.\ref{fig_1MT ramp1} for four different values of loading rate $a$ is, thus, an unambiguous evidence in favour of the catch-bond-like behavior exhibited by the kt-MT attachment (for $N=1$) also in our force-ramp experiment {\it in-silico}. Several different molecular mechanisms proposed so far can account for the observed signatures of catch bond in conventional ligand-receptor systems \cite{thomas08a,thomas08b,sokurenko08,prezhdo09,chakrabarti17}. However, the distinct mechanism that we have summarized above in the context of force-clamp studies of kt-MT attachment is responsible also for the catch-bond-like behavior displayed in the  Figs.\ref{fig_1MT ramp1} and \ref{fig_1MT ramp2}.  

In principle, our theoretical predictions for $N=1$ can be tested using the reconstituted kinetochore of budding yeast {\it in-vitro} \cite{akiyoshi12} applying standard techniques of dynamic force spectroscopy \cite{bizzarri12}; a typical set up would use an optical trap with controlled ramp protocol \cite{franck10}. In the force-clamp set up with optical trap, the bead-trap separation is maintained at a fixed value  with a computer controlled feedback while the change in the length of the MT is recorded by  monitoring the movement of the specimen stage \cite{franck10}. A force-ramp set up, where the  bead-trap separation is gradually increased with time, has also been designed by modifying the 
force-clamp software \cite{franck10}. This force-ramp can be used to test the corresponding  theoretical predictions made in this paper. 
However, the slow loading rate required to observe the theoretically predicted behavior may still pose technical challenges.

\section{Extended SSC model of MT- single kt attachment for N $>$ 1}

In this section we extend the SSC model to capture some key features of the energetics and kinetics of 
a dynamic attachment formed between a single kt and a bundle of $N$ parallel MTs. 
As mentioned in the introduction, this extension is motivated by the fact that, in almost all organisms, 
except for budding yeast, each kt can normally attach to multiple MTs simultaneously. However, in 
none of the organisms, other than budding yeast, the Dam1 ring, or any analogous complete ring-like 
structure, have been detected so far. Therefore, kt-MT coupling based on a real complete sleeve or 
ring seem highly unlikely in these systems \cite{mcintosh13}. 

\begin{figure}[htb]
\center
\includegraphics[angle=0,width=0.95\columnwidth]{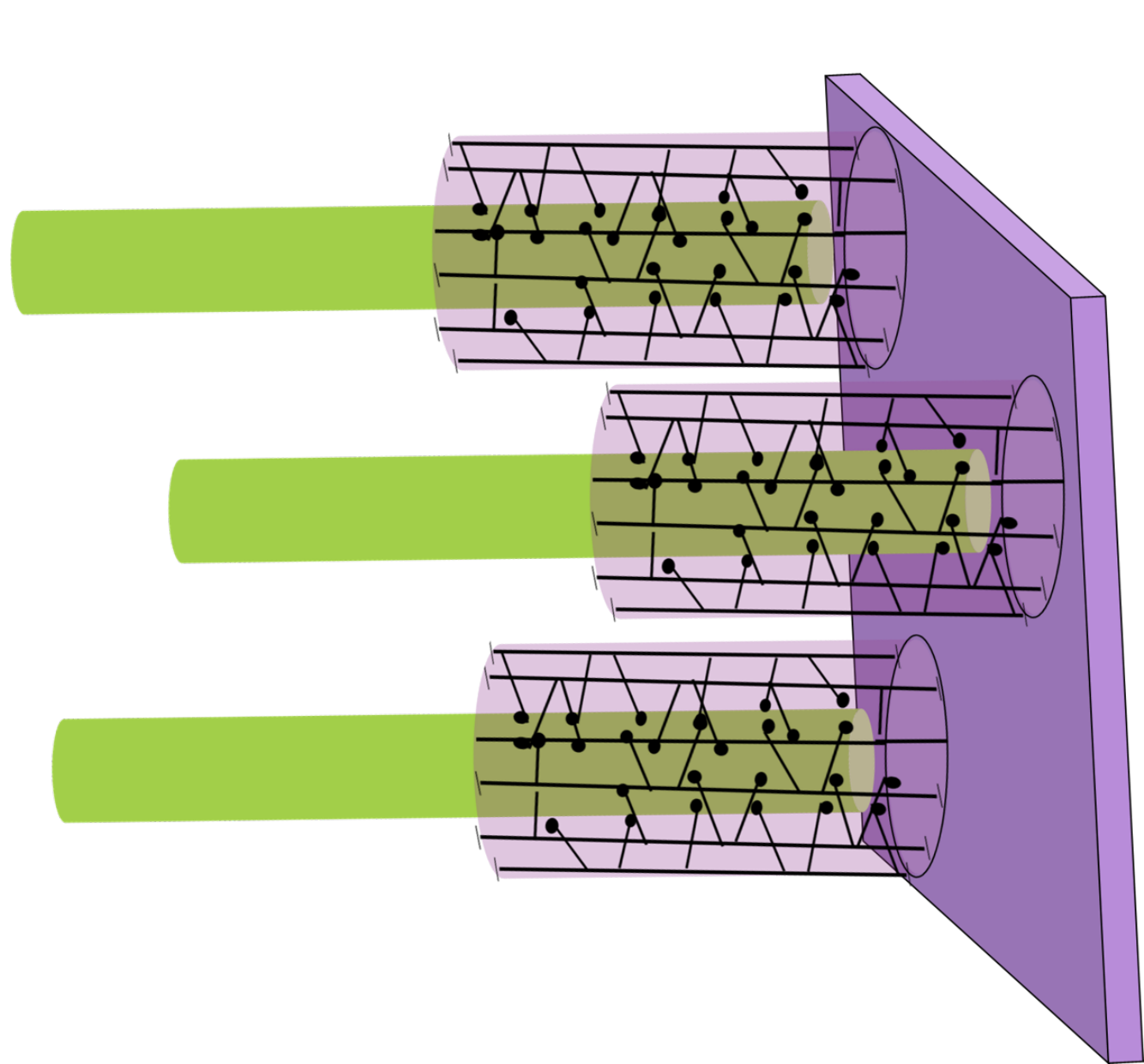}
\caption{Three microtubules (green cylinders) are attached to a single kinetochore (violet wall) in the presence of external tension on kinetochore (inspired by Fig.1(b) of ref.\cite{keener14}).} 
\label{fig_ktmt}
\end{figure} 

 
Based on the ultrastructure of vertebrate kinetochores, \cite{dong07,mcewen10,mcintosh08} 
and {\it in-vitro} molecular force spectroscopy \cite{powers09}, 
it is widely believed that flexible filamentous MT-binding proteins 
\cite{shtylla11,keener14}, that are components of a kinetochore, can form load-bearing attachments with MTs. 
The `binders' appear as one of the core concepts in several recent models that include also  the ``lawn'' model 
\cite{zaytsev14}, ``sliding foot'' model \cite{janczyk17} , etc. 
These binders can engage a MT from all angles (see Fig.\ref{fig_ktmt}). Moreover, unlike the synchronous attachments and detachments of the postulated MT-binding sites on the inner surface of Hill'��s sleeve \cite{hill85}, the attachment and detachment of these flexible filamentous binders are, in general, not synchronous. Furthermore, these filaments do not link among themselves permanently to form any rigid ring-like or sleeve-like structure.

Nevertheless, based on the observations in their {\it in-vitro} experiments and Monte Carlo simulations, Powers et al. \cite{powers09} argue that an effectively biased diffusion mechanism, similar to that postulated by Hill \cite{hill85}, can still emerge from the fibrous kt-MT linkers even if no rigid sleeve-like structure exist at the surface of a kinetochore. Therefore, the effective potential landscape has also been speculated \cite{santaguida09} to be 
qualitatively similar to that in the Hill sleeve model. Because of the possibility that the binders engage the MT surface practically uniformly and because of the finite maximum stretchable length of the binders, we {\it assume} that an effective sleeve-like region may be created  (see Fig.\ref{fig_ktmt}).

It is worth pointing out that the effective potential in the Hill sleeve model is corrugated because movement of the sleeve along the MT requires breaking and subsequent re-establishment of the bonds between MT-binding sites on the inner surface of the sleeve and their specific binding sites on the outer surface of the MT. In the simplest version of the SSC model used earlier in this paper, only the tilt of the corrugated potential was retained by assuming a linear potential energy landscape; the corrugation, which manifests as `molecular friction', was ignored. Even this simplified potential energy landscape was found to be adequate to get a deep insight into the physical mechanism of the catch-bond-like behaviour of the kt-MT attachment. 

In the same spirit, the effective potential energy landscape for every individual kt-MT attachment is {\it assumed} here also to be linear. Even during a period when $y$ remains fixed individual binders can attach to- or detach from the MT surface. Consequently, unlike the original Hill-sleeve model, a major component of the force pulling the MT towards the kt surface could be of entropic origin \cite{zaytsev13,zaytsev14}. A kt-ward pull exerted by a binder bound to 
curled protofilament at the tip of a depolymerizing MT can suppress the curling, and hence the rate of 
depolymerization of the MT just as the Dam1 ring does in case of budding yeast. 
Thus, both the two {\it postulates} (a) and (b), encapsulated by the eqns.(\ref{eq_Vb}) and (\ref{eq_betaF}), 
respectively, are {\it assumed} to remain valid for each individual MT, provided $V_{b}$ is interpreted as a potential of mean force.

We study the collective strength and stability of this attachment formed by a bundle of parallel MTs by 
computer simulation of molecular force spectroscopy under both force-clamp and force-ramp conditions.  
To our knowledge, no experimental data are available at present to make direct comparison with the predictions of the general model ($N > 1$) analyzed in this section. However, very recent experimental breakthroughs \cite{weir16}  suggest that both force-clamp and force-ramp experiments with reconstituted mammalian kinetochores {\it in-vitro} may become possible in near future.

In this extended SSC model at any arbitrary instant of time $t$, a single kt is attached to $n(t)$ ($1 \leq n(t) \leq N$) parallel MTs, each through its respective coupler, where $N$ is the maximum number of MTs that can attach to the kt simultaneously. For simplicity, all the couplers are assumed to have identical length $L$. The MTs are not directly coupled by any lateral bond (transverse to their axis). Instead, all the collective effects arise from their indirect coupling via the kinetochore to which $n(t)$ MTs are attached.  The physically motivated assumption of the model, which couple their kinetics is that at any instant of time $t$, the externally applied load tension $F$ is shared equally among the $n(t)$ MTs that are attached to the kt at that instant through their respective couplers, i.e., $F/n(t)$.

We consider two possible scenarios for the rupture of a joint formed by a kt initially with multiple MTs. In the {\it first}, once a MT detaches, its re-attachment to the same kt is not allowed. Number of MTs attached with kt, starting from the initial maximum value $N$, varies irreversibly as
\begin{equation}
N\rightarrow N-1 \rightarrow N-2 \rightarrow  N-3 \rightarrow...............2 \rightarrow 1 \rightarrow 0
\end{equation}
In the {\it second} scenario, once a MT detaches it can reattach again to the same kt and can grow inside the coupler because of its polymerization. So, in this case, the number of MTs attached to the kt varies reversibly as 
\begin{equation}
N\rightleftharpoons N-1 \rightleftharpoons N-2 \rightleftharpoons N-3 \rightleftharpoons...............2 \rightleftharpoons 1 \rightarrow 0
\end{equation}
where the last step is irreversible because of the absorbing boundary condition imposed at $n=0$.

\begin{figure}[htb]

\includegraphics[angle=-0,width=0.4\textwidth]{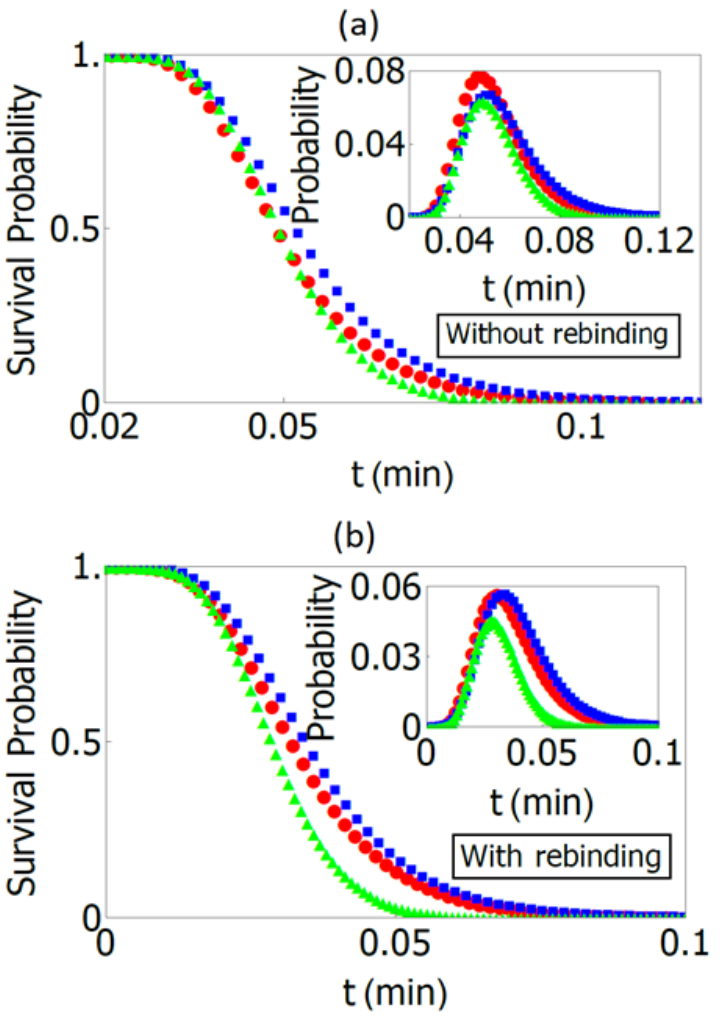}
\caption{
Survival probability is plotted as a function of time $t$, under force clamp condition (a) in the absence of rebinding, for three different values of the tension, $ F=0.01 $ pN, (red circle), $ F=0.5 $ pN (blue square) and $ F=1 $ pN (green triangle), and (b) in the presence of rebinding, for three values of the tension  $ F=0.01 $ pN (red circle), $ F=0.6 $ pN (blue square) and $ F=1.5 $ pN (green triangle). In the insets of both the figures the corresponding distributions of the lifetimes are shown. The numerical values of all the other parameters used in the simulation are listed in the table \ref{table-parameter} except, $ N=40 $, {\bf $ F_{*}=1 $} pN and $ B=1 $ pN.}
\label{fig_NMT_clamp1}
\end{figure} 

\begin{figure}[htb]

\includegraphics[angle=-0,width=0.4\textwidth]{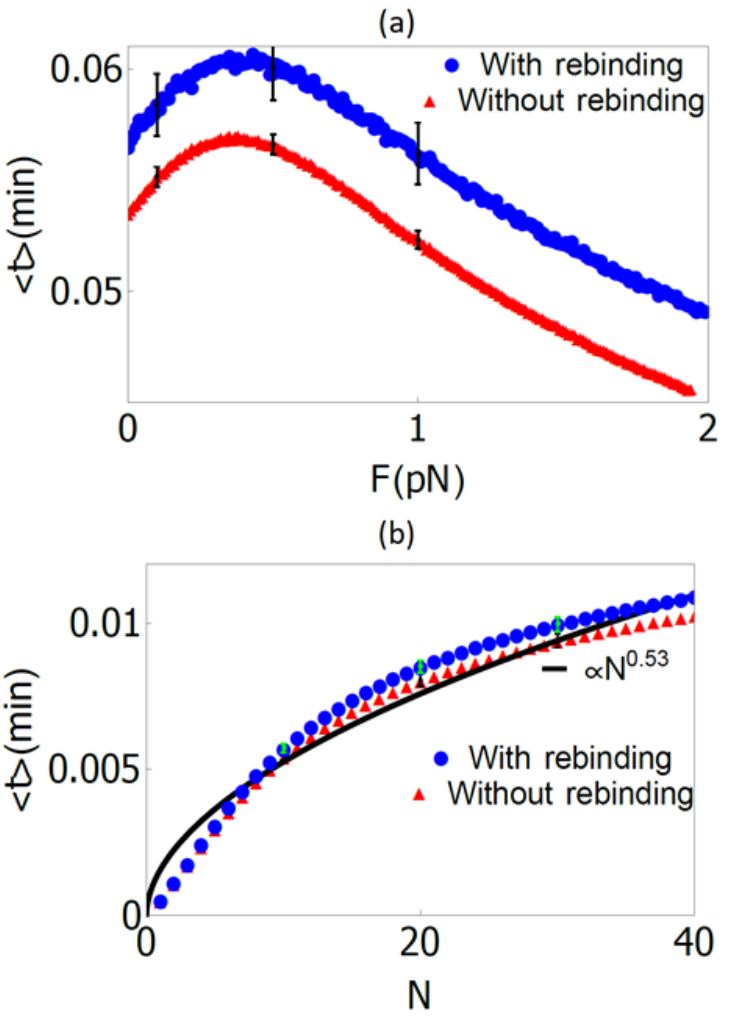}
\caption{
The mean life time $<t>$ is plotted against (a) tension $F$ for $N=40$ and (b) number $N$ for $F=10$ pN, $B=0.5$ pN, each for both the scenarios, namely with rebinding  (blue circle)  and without rebinding (red triangle). 
(b) We found best fit of our simulation data with the curve $ <t> \propto N^{0.53} $, represented by black continuous line.
The numerical values of all the other parameters used in the simulation are listed in the table \ref{table-parameter} except, $ N=40 $, {\bf $ F_{*}=1 $} pN and $ B=1 $pN. Error bars represent standard deviation of the simulation data.
}
\label{fig_NMT_clamp2}
\end{figure} 

\begin{figure}[htb]

\includegraphics[angle=-0,width=0.4\textwidth]{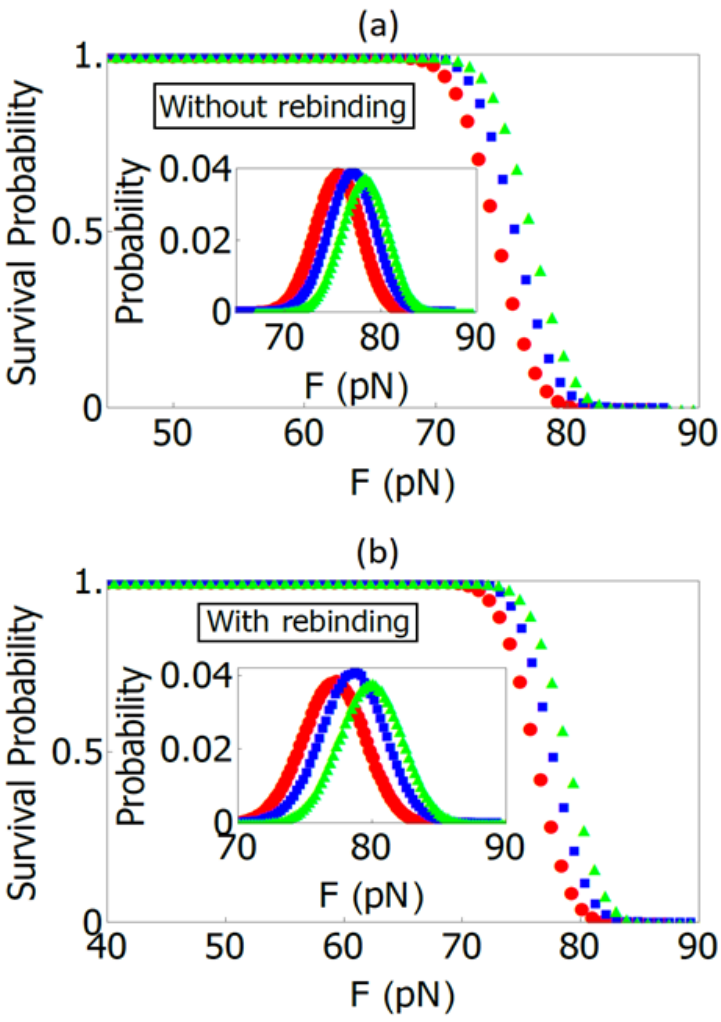}
\caption{
Survival probabilities, under force-ramp condition, for three different loading rates $ a= 18 $pN/s (red circle), $ 20 $pN/s (blue square) and $ 22 $pN/s (green triangle) are plotted (a) in the absence of rebinding and (b) in the presence of rebinding  . In the inset the corresponding distributions of the rupture forces are shown.  The numerical values of all the other parameters are listed in the table \ref{table-parameter}.}
\label{fig_NMT_ramp1}
\end{figure} 

Extending the WPE prescription \cite{wang03,wang07} used earlier for the single MT-kt attachment, space is now discretized into $ M $ cells, each of length $ h=L/M $. Then the time-dependent discrete effective potential is given by
\begin{equation}
   \frac{U_{nj}}{k_{B}T}=\biggl[\frac{\biggl(\frac{F}{n(t)}-B\biggr)}{k_{B}T}+{\ell}\frac{\beta_{0}e^{-\frac{F/n(t)}{F_{\star}}} -\alpha }{D}\biggr]y_{j}
 \label{eq_potential}
\end{equation}
where $n(t)$ is the number of MTs attached to the kt at the instant of time $t$.
Accordingly, the corresponding forward ($ w_{fn}(j)) $ and backward ($ w_{bn}(j)) $ transition rates  can be written by substituting $U_{nj}$ in the place of $\tilde{U_{j}} $ in the Eqns.(\ref{eq_wf}),(\ref{eq_wb}).
In our simulation of both the scenarios mentioned above, initially, all the $ N $ MTs are fully inserted into the kt coupler. 

In the first scenario, using the transition rates given by $ w_{fn}(j) $ and $ w_{bn}(j) $, the position of a MT tip inside its coupler is updated. But, once an attachment ruptures, its reattachment to the kt is not allowed; therefore, detached MT is no longer monitored in our simulation. However, the simulation is continued till the last surviving MT-kt attachment ruptures. This first passage time is identified as the life time of the molecular joint consisting of $N$ MTs with a single kt. The process is repeated $ 10^{6} $ times, starting from the same initial condition, to obtain the distribution of the lifetimes. In the same scenario, under the force-ramp condition ($ F=at $) we collect the data similarly to obtain the distribution of rupture forces (i.e., the force at which the tip of the last surviving MT exits from its coupler).
 
In the alternative scenario, the transition rates $ w_{fn}(j) $ and $ w_{bn}(j) $ govern the kinetics of the tip of each MT as long as it moves inside the corresponding coupler. However, once the attachment between a MT and the kt, through the coupler, ruptures it must get an opportunity to reattach through its natural kinetics of polymerization and depolymerization outside the coupler. Therefore, in this scenario, the continuing forward and backward movement of the tip of a detached MT outside its coupler is monitored in our simulation. During this period the force-free kinetics of the MT tip outside its coupler is implemented in our simulation by replacing the potential (\ref{eq_potential}) by the simpler potential
\begin{equation}
   \frac{V_{j}}{k_{B}T}={\ell} \biggl[\frac{\beta_{0} -\alpha }{D}\biggr]y_{j}
 \label{eq_potentialR}
\end{equation}
and simultaneously replacing the transition rates $ w_{fn}(j) $ and $ w_{bn}(j) $ by
\begin{equation}
 w_{f1}(j)=\frac{D}{h^{2}}\frac{-\frac{\delta V{j}}{k_{B}T}}{exp(-\frac{\delta V_{j}}{k_{B}T}-1)}
 \label{eq_wf11}
\end{equation}
and
\begin{equation}
 w_{b1}(j)=\frac{D}{h^{2}}\frac{\frac{\delta V_{j}}{k_{B}T}}{exp(\frac{\delta V_{j}}{k_{B}T}-1)}, 
 \label{eq_wb11}
\end{equation}
respectively, where $ \delta V_{j}=V_{j+1}-V_{j} $. If, through this kinetics outside the coupler, a MT succeeds in re-entering its coupler its kinetics reverts back to that governed by the transition rates $ w_{fn}(j) $ and $ w_{bn}(j) $. 
Thus, starting from the initial state the time evolution of all the MTs are monitored till the instant when, for the first time, none of the MTs is attached to the kt; this first-passage time is identified as the lifetime of the attachment.
Repeating this process we have obtained the distributions of the lifetimes in the second scenario. Similarly for the ramp force we have obtained the distribution of the rupture force which is defined as the force at which, for the first time, none of the MTs is attached to the kt.

\subsection{Results on life time distribution under clamp force for N $>$ 1}

In Fig.\ref{fig_NMT_clamp1} (a) and (b) survival probabilities of an attachment, consisting initially of $40$ MTs and a single kt, have been plotted as a function of time for the two cases where rebinding is (a) forbidden and (b) allowed, 
respectively. The attachment survives for longer duration in intermediate range of the clamp force ($ F=0.5$pN, blue square) than at the high and low strength of the tension. In the inset the corresponding distributions of the lifetimes of the attachments are also shown. 

The trends of variation of the survival probability with the clamp force indicates a catch-bond-like behavior. Indeed, this catch-bond-like behavior can be seen directly in Fig.\ref{fig_NMT_clamp2}(a) where the mean life time $<t>$, plotted  against the clamp force $F$, displays a maximum at a non-zero finite value of $F$ irrespective of whether rebinding of the MTs is allowed or forbidden.  The physical cause of the catch-bond-like behavior is the same as that pointed out in the special case $N=1$. Moreover, as expected on physical grounds, for any given $F$, the mean life time $<t>$ is higher if rebinding is allowed as compared to the mean life time in the absence of rebinding. 
 
In the Fig.\ref{fig_NMT_clamp2}(b) mean lifetime is found to increase with the number of microtubule (N). This is consistent with one's intuitive expectation.  Besides, for any given value of $N$, allowing rebinding of the MTs results in a higher life time. However, the interesting point is that the mean life time does not exhibit trivial linear increase with $N$. Instead, it increases nonlinearly (more precisely, sub-linearly) with $N$ in both the cases. Though the parallel MTs do not interact with one another laterally but only by equal sharing of the instantaneous load, the nonlinear behavior is a collective emergent property of the interacting system. Recent reconstitution of mammalian kt {\it in-vitro} \cite{weir16} have raised the hope of indicates promising new routes for testing our results for $N > 1$.

\subsection{Results on rupture force distribution under force ramp for N $>$ 1}

\begin{widetext}

\begin{figure}[htb]
\includegraphics[angle=-0,width=0.8\columnwidth]{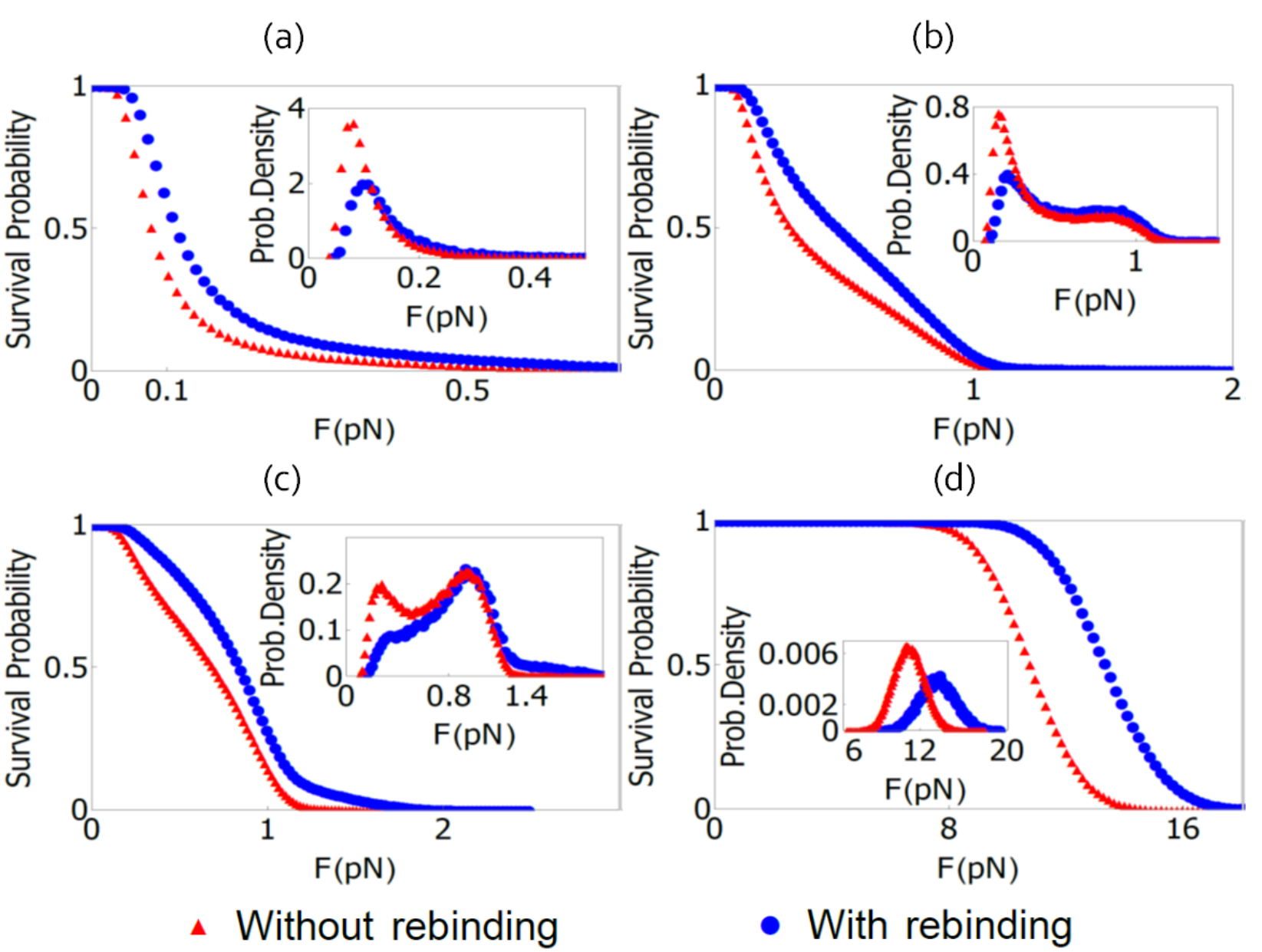}\\
\caption{Survival probability under force-ramp condition and probability density of rupture force (in the inset) of the kt-MT attachment with N = 40 for both in the presence and absence of rebinding for four different loading rates, namely, (a) $a =1\times 10^{-2} pN s^{-1}$, (b) $a = 2\times 10^{-2} pN s^{-1}$, (c) $a = 3\times 10^{-2} pN s^{-1}$ and (d) $a = 0.1 pN s^{-1}$ are plotted. The numerical values of all the other parameters used in the simulation are listed in the table \ref{table-parameter} except, $N=40$, $ F_{*}=0.5 $pN, $ B=1.5 $pN and $ \alpha=50 s^{-1}$.} 
\label{fig_NMT_ramp3}
\end{figure}

\end{widetext}

\begin{figure}[htb]
\includegraphics[angle=-0,width=0.4\textwidth]{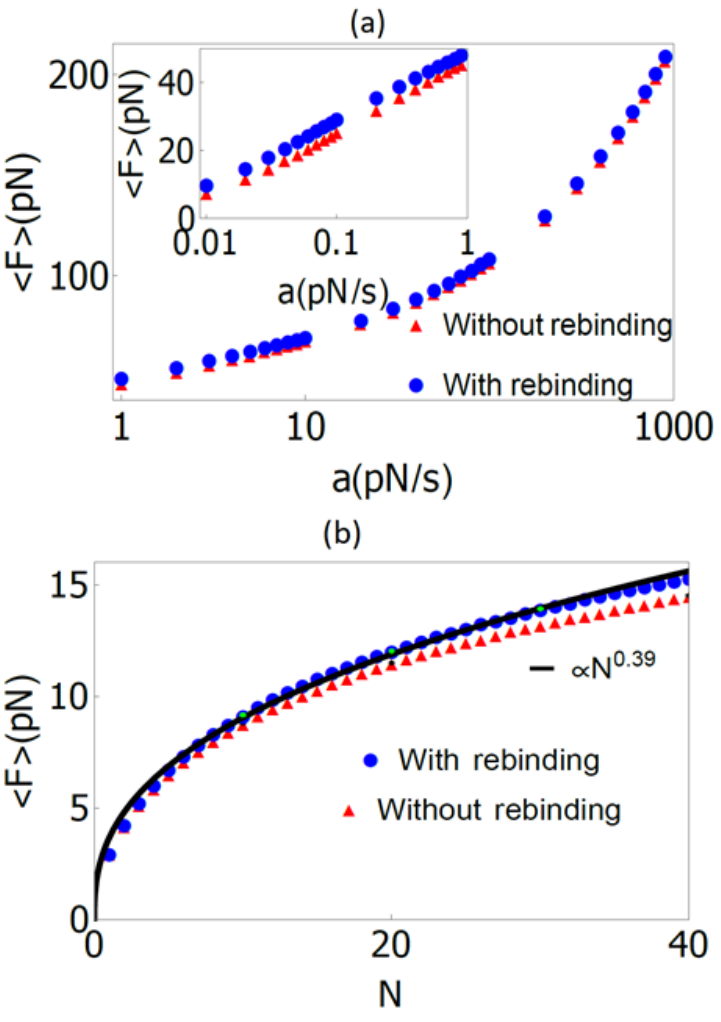}
\caption{
 Mean rupture force is plotted against (a) the loading rate $a$ for a fixed $N=40$, and (b) $N$ for a fixed loading rate $a=10$ pN/s,  $ F_{*}=1 $ pN and $ B=1 $ pN. Logarithmic scale is used along the X-axis in (a) to cover a very broad range of $a$. In the inset of (a) the mean rupture force is plotted for relatively lower loading rates where the difference in the data for the two cases, namely with and without rebinding, are more pronounced and clearly visible. {\textbf (b) The  best fit to our simulation data is obtained with the curve  $ <F> \propto N^{0.39} $, represented by black continuous line.
Error bars (small green dot) represent standard deviation.}
 The numerical values of all the other parameters are listed in the table \ref{table-parameter}.}
\label{fig_NMT_ramp2}
\end{figure} 

In the Fig.\ref{fig_NMT_ramp1}(a) and (b)  survival probabilities (and the corresponding rupture force distribution in the insets) are plotted, respectively,  in the absence and presence of rebinding for three different loading rates $ a=18$pN/s, $20$pN/s and $ 22 $pN/s. Survival probability remains high upto a certain force beyond which it drops quite sharply. 
The rupture force distribution in this figure does not display the bimodal form seen earlier in Fig. \ref{fig_1MT ramp1} and Fig. \ref{fig_1MT ramp2} for N=1. In contrast, in the Fig.\ref{fig_NMT_ramp3}, where the survival probabilities and the corresponding rupture force distribution (in the insets) are plotted for a slightly different set of values of the key parameters $F_{\ast}$ and $B$, a bimodal form is found. Moreover, the trend of variation of rupture force distribution and survival probability is similar to those observed in Fig. \ref{fig_1MT ramp1} and Fig. \ref{fig_1MT ramp2}(a) for N=1 scenario. There is a minor difference between the bimodal forms of the rupture force distributions in Fig.\ref{fig_1MT ramp1} for N=1 and Fig.\ref{fig_NMT_ramp3} for $N > 1$; the first peak in the former appears at $F=0$ whereas that in the latter corresponds to a non-zero value of $F$. However, both are consistent with earlier reports on different ligand-receptor bonds \cite{evans04,thomas08a,thomas08b,kim10}.
The contrast of the qualitative trends of variation of the rupture force distributions in the Figs.\ref{fig_NMT_ramp1} and \ref{fig_NMT_ramp3} also emphasizes the role of the importance of the energetics and kinetics of the MTs in the catch-bond-like behavior.

In the Fig.\ref{fig_NMT_ramp2}(a) and (b) the average rupture force is plotted, respectively, against the loading rate  $a$ (for a given $N$) and against $N$ (for a given loading rate $a$). The log-scale along the X-axis in Fig.\ref{fig_NMT_ramp2}(a) is used to cover a wide range of loading rates in the most suitable manner. The higher survival probability caused by reattachment of MTs is more pronounced at slower loading than at faster loading. This trend of variation follows from the fact that at faster loading detached MTs get smaller chances of reattaching before the complete rupture of the attachment.  What is interesting from the quantitative point of view is that the average rupture force increases nonlinearly with increasing loading rate.
For high loading rate average rupture force $ <F>$ follows a linear trend \cite{williams03}, but here nonlinear behavior arises because, faster loading rates allow less time for the dissociation and depolymerization processes, ultimately leading to rupture of MT-kt bonds. Finally, the increase of the mean rupture force with increasing $N$ also seems to be nonlinear.

\section{Discussion, Summary and conclusion}

In this paper we have developed theoretical models of molecular joints formed by $N (> 1)$ parallel MT filaments with a single kt by extending the SSC model \cite{sharma14} that was developed for the special case $N=1$. 
By carrying out extensive kinetic Monte Carlo simulations of our theoretical models of kt-MT attachments, we have computed the probability distributions, and hence the mean values, of the lifetimes and rupture forces which are the two main characteristic statistical properties of such transient attachments.

The SSC model with $N=1$ \cite{sharma14} not only reproduced the catch-bond-like behaviour of the kt-MT attachments  observed in force-clamp experiments {\it in-vitro} for budding yeast \cite{akiyoshi10}, but also elucidated a plausible underlying mechanism that gives rise to this counter-intuitive phenomenon. However, 
in ref.\cite{sharma14}, the lifetimes of the attachments were calculated for only a single unique initial condition. 
In the first half of this paper we have presented new results for some other initial conditions to convincingly 
establish that the qualitative conclusions drawn in ref.\cite{sharma14} are valid for all possible initial conditions. 
Moreover, we have presented further evidence in favour of the catch-bond-like behavior, for the same $N=1$ 
case, by reporting `mechanical signatures' of typical catch-bond observed in in our {\it in-silico} force-ramp experiments.

In the second half of this paper we have extended the SSC model to the more general case $N > 1$. In this case, 
the possibility of re-attachment of a detached MT to the same kt, before the last surviving MT gets detached, can prolong the lifetime. We present simulation data to establish that, in spite of this additional complexity that did not exist in the special case $N=1$, the kt-MT attachment still exhibits a catch-bond-like behavior in a part of the parameter space of this model.
As a byproduct of this investigation we also find that both the mean lifetime and mean rupture force scale nonlinearly with $N$. This result is important from the perspective of collective phenomena. Although in our models there is no direct lateral interaction among the MTs, the indirect interactions among the MTs is mediated by the kt to which all the MTs are attached. This indirect interactions give rise to the non-trivial nonlinear scaling of the mean lifetime and mean rupture force with $N$. Similar trends of variation of non-covalent bond rupture characteristics with increasing number of ligands have been observed in the past \cite{sulchek05}. 

The SSC model \cite{sharma14}, and its extensions reported in this paper, are minimal models based on two key assumptions encapsulated by the Eqs.(\ref{eq_Vb}) and (\ref{eq_betaF}). The first postulate (\ref{eq_Vb}) incorporates the main feature of the energetics of MT-coupler interactions that implicitly depends also on the structure of the kt-MT coupler. The second postulate (\ref{eq_betaF}) captures the most essential aspect of the kinetics of depolymerization of microtubules under load tension. These minimal models draw heavily on biased diffusion of Hill's sleeve \cite{hill85} and  conformational wave  based on curling of depolymerizing tip of MT \cite{koshland88}. Both these models, however, were proposed long before the composition and structure of kt could be explored at the molecular level \cite{cheeseman14}. We have argued that our minimal models can also be interpreted so as to make these consistent with the recent structural models of mammalian kinetochores because our minimal models do not explicitly assume any specific structure of the kt-MT coupler.  A mechanical model, in terms of beads connected by springs, was developed by Bertalan et al.\cite{bertalan14};  in this model, the attachment is assumed to be formed by the insertion of the curling protofilament hook into the loops formed by the kinetochore fibrils. It has not been possible to identify measures that would differentiate between our kinetic models and the more explicit structural model developed by Bertalan et al. \cite{bertalan14}.

One of the unique features of the polymerization kinetics of a MT is its dynamic instability \cite{desai97}. A polymerizing MT keeps growing in length till it suffers a ``catastrophe'' whereby it abruptly begins to depolymerize. A depolymerizing MT would, eventually, disappear unless its rapid shrinkage is stopped by a process called `rescue'' following which it resumes polymerization. 
The theory for this phenomenon of dynamic instability, that began with Hill's pioneering work \cite{hill84}, has been re-formulated and improved over the subsequent years \cite{dogterom93,bicout97,antal07,sumedha11,needleman10,brugues12,ishihara16,decker18} (see ref.\cite{ataullakhanov13,anderson13,anderson15} for reviews).

The 2-state model that Akiyoshi et al.\cite{akiyoshi10} used to account for their experimental data explicitly describe switching of the MT between the growing and shrinking stages because of catastrophe and rescue. This model was extended by Zhang \cite{zhang11} assigning additional distinct mechano-chemical states that enable  capturing the dependence of the MT catastrophe rate on the GTP-tubulin concentration. However, neither of these two versions of the 2-state model throw light on the physical origin of the phenomenon in terms of the structure and dynamics of the kt-MT attachment. 
Any explicit description of the kinetics of the growing and shrinking MTs separately would require equations that govern the time evolutions of probability densities  $P_{\pm}(x,t)$ of the polymerizing (+) and depolymerizing (-) MTs. In contrast, the SSC model, as well as the extended versions studied in this paper, describe MT kinetics in terms of a single probability density $P(x,t) = P_{+}(x,t) + P_{-}(x,t)$. The assumption that $P(x,t)$ alone provides an alternative, but adequate, description of the generic features of the molecular force spectroscopy of kt-MT attachment is an assumption that is well justified by the results.

In recent years, strain-dependent detachment of molecular motors like dynein and myosin have revealed catch-bond like stabilization of the track-bound state of the motor by externally applied tension \cite{erdmann12,erdmann16,inoue16,nair17}. 
Such catch-bonds have important biological functions in cell adhesion, mechanosensation, mechano-transduction, immune response, bacterial mechanics, etc. \cite{vernerey16,leckband14,huse17,persat15}. Here we have modelled and analyzed the kt-MT attachment by drawing analogy with common ligand-receptor bonds. Elsewhere we have invoked similar analogies \cite{ghanti17,chowdhury18} for studying transient attachments formed by MTs in the mitotic spindle, namely at the cell cortex \cite{wu17} and at the spindle pole \cite{fong17}. 
Conceptually, this a leap forward because the MTs, the analogs of ligands,  are self-organized supra-molecular structures made of building blocks each of which itself is a macromolecule while the kt, the counterpart of a receptor, is also a complex structure made of made macromolecules. 

In case of common ligands at least three different geometries can be distinguished; (a) $N$ ligands in parallel where each one is subjected to a load $F/N$ if the load $F$ is shared equally by all, (b) $N$ ligands in series where all the ligands are subjected to the same load $F$, and (c) $N$ ligands in `zipper' configuration where only the bond at the leading edge bears the entire load $F$ while no load is experienced by the others. Moreover, in case of parallel geometry, the flexibility of the long ligands can have significant effect on the manner in which the load is shared. In contrast,  each MT is quite stiff. Our model with $N>1$ corresponds to the `parallel' geometry where, at any instant,  the load is shared equally by those MTs that are still attached to the kt at that instant of time.  

We also stress that, in spite of these  superficial similarities, there are several crucial differences in the underlying physical mechanisms because of which none of  the mechanism responsible for the catch-bonds in common ligand-receptor systems  \cite{bargesov05,chakrabarti14,chakrabarti17,rakshit14,liu15} is directly applicable to the kt-MT attachment. The main sources of these differences arise from the fact that (i) each MT tip can grow or shrink because of ongoing polymerization or depolymerization of the MT and (ii) the rate of depolymerization is strongly suppressed by externally applied tension. 
It is precisely for this reason that we regard the kt-MT attachments as ``unusual''  in spite of the fact they display the usual signatures of catch-bonds.

\section*{ACKNOWLEDGEMENT}
One of the authors (DC) thanks Charles Asbury for valuable comments on a shorter preliminary draft of this manuscript. DC also thanks Raymond Friddle, Gaurav Arya, D. Thirumalai and Shaon Chakrabarti for useful correspondences. This work has been supported by a J.C. Bose National Fellowship (DC) and ``Prof. S. Sampath Chair'' Professorship (DC).

\section*{REFERENCES}


\begin{thebibliography}{99}

\bibitem{wittmann01} T. Wittmann, A. Hyman and A. Desai, Nat. Cell Biol. {\bf 3}, E28 (2001).

\bibitem{karsenti01} E. Karsenti and I. Vernos, Science {\bf 294}, 543 (2001).

\bibitem{helmke13} K.J. Helmke, R. Heald and J.D. Wilbur, Int. Rev. Cell Mol. Biol. {\bf 306}, 83 (2013).

\bibitem{karsenti08} E. Karsenti, Nat. Rev. Mol. Cell Biol. {\bf 9}, 255 (2008). 
\bibitem{frank11} J. Frank (ed.), {\it Molecular Machines: Workshops of the cell} (Cambridge University Press, 2011). 

\bibitem{mcintosh12} J.R. McIntosh, M.I. Molodotson and F.I. Ataullakhanov, Quart. Rev. Biophys. (2012).

\bibitem{dumont14} S. Dumont and M. Prakash, Mol. Biol. of the Cell {\bf 25}, 3461 (2014). 

\bibitem{reber15} S. Reber and A.A. Hyman, Cold Spring Harb. Perspct. Biol. {\bf 7}, a015784 (2015).

\bibitem{lawson13} J. L.D. Lawson and R.E. C. Salas, Biochem. Soc. Trans. {\bf 41}, 1736 (2013). 

\bibitem{cheeseman14} I.M. Cheeseman, Cold Spring Harb. Perspect. Biol. {\bf 6}, a015826 (2014). 

\bibitem{petry16} S. Petry, Annu. Rev. Biochem. {\bf 85}, 659 (2016).

\bibitem{yusko14} E.C. Yusko and C.L. Asbury, Mol. Biol. of the Cell {\bf 25}, 3717 (2014). 

\bibitem{margolis81} Margolis and Wilson, Nature (2981).

\bibitem{evans98} E. Evans, Faraday Discuss., {\bf 111}, 1 (1998).

\bibitem{karplus10} M. Karplus, J. Mol. Recognit. {\bf 23}, 102 (2010). 

\bibitem{bizzarri12} A.R. Bizzarri and S. Cannistraro (eds.) {\it Dynamic Force Spectroscopy and Biomolecular Recognition}, (CRC Press, 2012).

\bibitem{franck10} A.D. Franck, A.F. Powers, D.R. Gestaut, T.N. Davis and C.L. Asbury, Methods {\bf 51}, 242 (2010).

\bibitem{biggins13} S. Biggins, genetics {\bf 194}, 817 (2013). 

\bibitem{akiyoshi12} B. Akiyoshi and S. Biggins, Chromosoma {\bf 121}, 235 (2012).

\bibitem{sarangapani14} K.K. Sarangapani and C.L. Asbury, Trends in Genet. {\bf 30}, 150 (2014).

\bibitem{akiyoshi10} B. Akiyoshi, K. K. Sarangapani, A. F. Powers, C. R. Nelson, S. L. Reichow, H. Arellano-Santoyo, 
T. Gonen, J. A. Ranish, C. L. Asbury and S. Biggins, Nature {\bf 468}, 576 (2010). 

\bibitem{thomas08a} W. E. Thomas, Annu. Rev. Biomed. Eng. {\bf 10}, 39 (2008).

\bibitem{thomas08b} W. E. Thomas, V. Vogel and E. Sokurenko, Annu. Rev. Biophys. {\bf 37}, 399 (2008).

\bibitem{sokurenko08} E.V. Sokurenko, V. Vogel and W.E. Thomas, Cell Host \& Microbe {\bf 4}, 314 (2008)

\bibitem{prezhdo09} O.V. Prezhdo and Y.V. Pereverzev, Acc. Chem. Res. {\bf 42}, 693 (2009). 

\bibitem{chakrabarti17} S. Chakrabarti, M. Hinczewski and D. Thirumalai, J. Struct. Biol. {\bf 197}, 50 (2017).  

\bibitem{bargesov05} V. Bargeson, D. Thirumalai, PNAS {\bf 102}, 1835 (2005).

\bibitem{sharma14} A. K. Sharma, B. Shtylla and D. Chowdhury, Phys. Biol.{\bf 11}, 1478 (2014).

\bibitem{bameta17} T. Bameta, D. Das, D. Das, R. Padinhateeri and M. Inamdar, Phys. Rev. E {\bf 95}, 022406 
(2017).

\bibitem{hill85} T. Hill, Proc. Natl. Acad. Sci. U.S.A. {\bf 82}, 4404 (1985).

\bibitem {buttrick11} G.J. Buttrick and J.B.A. Millar, Chromosome Res. {\bf 19}, 393 (2011).

\bibitem{westermann07} S. Westermann, D.G. Drubin and G. Barnes, Annu. Rev. Biochem. {\bf 76}, 563 (2007).

\bibitem{davis07} T.N. Davis and L. Wordeman, Trends in Cell Biol. {\bf 17}, 377 (2007).

\bibitem{foley13} E.A. Foley and T.M. Kapoor, Nat. Rev. Mo. Cell Biol. {\bf 14}, 25 (2013). 

\bibitem{efremov07} A. Efremov, E.L. Grishchuk, J.R. McIntosh and F.I. Ataullakhanov, PNAS {\bf 104}, 19017 (2007).

\bibitem{asbury11} C.L. Asbury, J.F. Tien and T.N. Davis, Trends in Cell Biol. {\bf 21}, 38 (2011).

\bibitem{grishchuk17} E.L.  Grishchuk, in: {\it Centromeres and Kinetochores}, ed. B.E. Black (Springer, 2017). 

\bibitem{anderson13} H. Bowne-Anderson, M. Zanic, M. Kauer and J. Howard, Bioessays {\bf 35}, 452 (2013). 

\bibitem{franck07} A. D. Franck, A.F. Powers, D.R. Gestaut, T. Gonen, T. N. Davis and C.L. Asbury, Nat. Cell Biol. {\bf 9}, 832, (2007).

\bibitem{risken} H. Risken, {\it The Fokker-Planck Equation} (Springer, 1996).

\bibitem{mirny10} L.A. Mirny and D.J. Needleman, Meth. Cell Biol. {\bf 95}, 583 (2010). 

\bibitem{joglekar02} A. P. Joglekar and A.J. Hunt, Biophys. J. {\bf 83}, 42 (2002). 

\bibitem{shtylla11} B. Shtylla and J. P. Keener, SIAM J. Appl. Math. {\bf 71}, 1821 (2011).

\bibitem{waters96} J. C. Waters, T.J. Mitchison, C.L. Rieder and E. D. Salmon, Mol. Biol. Cell. {\bf 7}, 1547 (1996).

\bibitem {wang03} H. Wang, C. Peskin and T. Elston, J. Theo. Biol. {\bf 221}, 491 (2003).

\bibitem {wang07} H. Wang, T. C. Elston, J. Stat. Phys. {\bf 128}, 35 (2007).

\bibitem{bell78}G.I. Bell. Science, {\bf 200}, 618, (1978).

\bibitem{evans97} E. Evans and K. Ritchie, Biophy J.,{\bf 72}, 1541, (1997).

\bibitem{evans02} E Evan and P. Williams, in: {\it Phys. of Biomolecules and Cells}, eds. H. Flyvbjerg, F. J\"ulicher, F. Ormos and F. David (Springer and EDP Sciences, 2002).

\bibitem{evans01} E. Evans, Annu. Rev. Biophys. Biomol. Struct. {\bf 30}, 105 (2001). 

\bibitem{friddle12} R. W. Friddle, in: {\it Dynamic Force Spectroscopy and Biomolecular Recognition}, ed. A.R. Bizzarri and S. Cannistraro (CRC Press, 2012). 

\bibitem{arya16} G. Arya, Molec. Simul. {\bf 42}, 1102 (2016).

\bibitem{gonen12} S. Gonen, B. Akiyoshi, M.G. Iadanza, D. Shi, N. Duggan, S. Biggins, T. Gonen, Nat. Struct. Mol. Biol. {\bf 19}, 925 (2012).

\bibitem{bloom08} A. P. Joglekar, D. Bouck, K. Finley, X. Liu, Y. Wan, J. Berman, X. He, E.D. Salmon and K.S. Bloom, J. Cell Biol. {\bf 181}, 587 (2008).

\bibitem{johnson10} K. Johnston, A. Joglekar, T. Hori, A. Suzuki, T. Fukagawa, and E.D. Salmon, J. Cell Biol. {\bf 189}, 937 (2010).

\bibitem{salmon06} A. P. Joglekar, D. C. Bouck, J. N. Molk, K. S. Bloom and E. D. Salmon, Nat. Cell Biol. {\bf 8}, 581 (2006).

\bibitem{marshall01} W. F. Marshall, J. F. Marko, D. A. Agard and J. W. Sedat, Curr. Biol. {\bf 11}, 569 (2001). 

\bibitem{evans04} E. Evans, A. Leung, V. Heinrich and C. Zhu, PNAS {\bf 101}, 11281 (2004). 


\bibitem{mcintosh13} J.R. McIntosh, E. O'Toole, K. Zhudenkov, M. Morphew, C. Schwartz, F.I. Ataullakhanov and E.L. Grishchuk, J. Cell Biol. {\bf 200}, 459 (2013).

\bibitem{dong07} Y. Dong, K.J. VandenBeldt, X. Meng, A. Khodjakov and B.F. McEwen, Nat. Cell Biol. {\bf 9}, 516 (2007).

\bibitem{mcewen10} B.F. McEwen and Y. Dong, Cell. Mol. Life Sci. {\bf 67}, 2163 (2010). 

\bibitem{mcintosh08} J.R. McIntosh, E.L. Grishchuk, M.K. Morphew, A. K. Efremov, K. Zhudenkov, V.A. Volkov, I.M. Cheeseman, A. Desai, D.N. Mastronarde and F.I. Ataullahkhanov, Cell, {\bf 135}, 322 (2008). 

\bibitem{powers09} A.F. Powers, A.D. Franck, D.R. Gestaut, J. Cooper, B. Gracyzk, R.R. Wei, L. Wordeman, T.N. Davis and C.L. Asbury, Cell {\bf 136}, 865 (2009). 

\bibitem{santaguida09} S. Santaguida and A. Musacchio, EMBO J. {\bf 28}, 2511 (2009). 

\bibitem{keener14} J. P. Keener and B. Shtylla, Biophys. J. {\bf 106}, 998 (2014). 

\bibitem{zaytsev13} A.V. Zaytsev, F.I.  Ataullakhanov and E.L. Grishchuk, Cell. Mol. Bioengg. {\bf  6}, 393 (2013).

\bibitem{zaytsev14} A.V. Zaytsev, L.J.R. Sundin, K.F. DeLuca, E.L. Grishchuk and J.G. DeLuca, J. Cell Biol. {\bf 206}, 45 (2014).

\bibitem{janczyk17} P.L. Janczyk, K.A. Skorupka, J.G. Tooley, D.R. Matson, C.A. Kestner, T. West, O. Pornillos and P.T. Stuckelberg, Dev. Cell {\bf 41}, 438 (2017).

\bibitem{weir16} J.R. Weir et al. Nature {\bf 537}, 249 (2016).

\bibitem{kim10} J. Kim, C.Z. Zhang, X. Zhang and T.A. Springer, Nature {\bf 466}, 992 (2010).

\bibitem{williams03} P. Williams, Anal. Chim. Acta.,{\bf 479}, 107, (2003).

\bibitem{sulchek05} T.A. Sulchek, R.W. Friddle, K. Langry, E.Y. Lau, H. Albrecht, T.V. Rattp, S.J. DeNardo, M.E. Colvin and Al. Noy, PNAS {\bf 102}, 16638 (2005).

\bibitem{koshland88} D.E. Koshland, T.J. Mitchison and M.W. Kirschner, Nature {\bf 331}, 499 (1988).


\bibitem{bertalan14} Z. Bertalan, C.A.M. La Porta, H. Maiato and S. Zapperi, Biophys. J. {\bf 107}, 289 (2014).

\bibitem{desai97} A. Desai and T.J. Mitchison, Annu. Rev. Cell Dev. Biol. {\bf 13}, 83 (1997).

\bibitem{hill84} T.L. Hill, Proc. Natl. Acad. Sci. USA {\bf 81}, 6728 (1984).

\bibitem{dogterom93} M. Dogterom and S. Leibler, Phys Rev Lett. {\bf 70}, 1347 (1993). 

\bibitem{bicout97} D. J. Bicout, Phys. Rev. E. {\bf 56}, 6656-6667, (1997). 

\bibitem{antal07} T. Antal, P.L. Krapivsky, S. Redner, M. Mailman and B. Chakraborty, Phys. Rev. E {\bf 76}, 041907 (2007). 

\bibitem{sumedha11} Sumedha, M.F. Hagan and B. Chakraborty, Phys. Rev. E {\bf 83}, 051904 (2011). 

\bibitem{needleman10} D.J. Needleman, A. Gronen, R. Ohi, T. Maresca, L. Mirny and T. Mitchison, Mol. Biol. Cell {\bf 21}, 323 (2010).

\bibitem{brugues12} J. Brugues, V. Nuzzo, E. Mazur and D.J. Needleman, Cell {\bf 149}, 554 (2012). 

\bibitem{ishihara16} K. Ishihara, K.S. Korolev and T.J. Mitchison, eLife {\bf 5}, e19145 (2016). 

\bibitem{decker18} F. Decker, D. Oriola, B. Dalton and J. Brugues,  preprint (2018). 

\bibitem{ataullakhanov13} F.I. Ataullakhanov, K.S. Melnik and A.A. Butylin, Biophys. (Moscow), {\bf 58}, 120 (2013). 

\bibitem{anderson15} H. Bowne-Anderson, A. Hibbel and J. Howard, Trends Cell Biol. {\bf 25}, 769 (2015).

\bibitem{zhang11} Y. Zhang, J. Biol. Chem. {\bf 286}, 39439 (2011). 

\bibitem{erdmann12} T.. Erdmann and U.S. Schwarz, Phys. Rev. Lett. {\bf 108}, 188101 (2012). 

\bibitem{erdmann16} T. Erdmann , K. Bartelheimer and U.S. Schwarz, Phys. Rev. E {\bf 94}, 052403 (2016).

\bibitem{inoue16} Y. Inoue and T. Adachi, Phys. Rev. E {\bf 93}, 042403 (2016). 

\bibitem{nair17} A. Nair, S. Chandel, M. Mitra, S. Muhuri and A. Chaudhuri, Phys. Rev. E {\bf 94}, 032403 (2016).

\bibitem{vernerey16} F.J. Vernerey and U. Akalp, Phys. Rev. E {\bf 94}, 012403 (2016).  

\bibitem{leckband14} D.E. Leckband and J. de Rooij, Annu Rev. Cell Dev. Biol. {\bf 30} 291 (2014). 

\bibitem{huse17} M. Huse, Nat. Rev. Immunol. {\bf 17}, 679 (2017).

\bibitem{persat15} A. Persat et al. Cell {\bf 161}, 988 (2015). 

\bibitem{ghanti17} D. Ghanti, R.W. Friddle and D. Chowdhury, preprint (2017). 

\bibitem{chowdhury18} D. Chowdhury et al. (2018). 

\bibitem{wu17} H.Y. Wu, E. Nazockdast, M.J. Shelley and D.J. Needleman, Bioessays {\bf 39},  1600212 (2016).

\bibitem{fong17} K. K. Fong, K.K. Sarangapani, E.C. Yusko, M. Riffle, A. Llauro, T.N. Davis and C.L. Asbury, Mol. Biol. Cell (2017).

\bibitem{chakrabarti14} S. Chakrabarti, M. Hinczewski and D. Thirumalai, PNAS {\bf 111}, 9048 (2014). 

\bibitem{rakshit14} S. Rakshit and S. Sivasankar, Phys. Chem. Chem. Phys. {\bf 16}, 2211 (2014). 

\bibitem{liu15} B. Liu, W. Chen and C. Zhu, Annu. Rev. Phys. Chem. {\bf 66}, 427 (2015). 

\end{thebibliography}
\end{document}